\documentclass[amsmath,amssymb,aps,pra,twocolumn,superscriptaddress]{revtex4-2}

\usepackage{amsmath}
\usepackage{mathtools}
\usepackage{physics}
\usepackage{dsfont}
\usepackage{setspace}
\usepackage{xcolor}
\usepackage{graphicx}
\usepackage[percent]{overpic}
\usepackage{svg}
\usepackage{textgreek}
\usepackage{amssymb}
\usepackage{comment}
\usepackage{subcaption}
\usepackage{tabularx}
\usepackage[labelformat=simple]{subcaption}
\usepackage{float}
\usepackage{hyperref}
\usepackage{titlesec}
\usepackage[toc,page]{appendix}

\begin{document}
\title{Scalable framework for quantum transport across large physical networks}
\author{Adam Burgess}%
\affiliation{SUPA, Institute of Photonics and Quantum Sciences, Heriot-Watt University, Edinburgh, EH14 4AS, UK}
\author{Nicholas Werren}%
\affiliation{SUPA, Institute of Photonics and Quantum Sciences, Heriot-Watt University, Edinburgh, EH14 4AS, UK}

\author{Erik M.~Gauger}%
\affiliation{SUPA, Institute of Photonics and Quantum Sciences, Heriot-Watt University, Edinburgh, EH14 4AS, UK}

\begin{abstract}
Accurately modelling many-body quantum transport systems poses a challenge both conceptually and computationally due to the growth of the Hilbert space and the multi-scale nature of the geometries and couplings present in most naturally occurring networks. A compounding complexity of such systems is that the environment typically plays a key role in the transport dynamics. Utilising variational unitary transformations that displace environmental degrees of freedom allows for the deployment of a second-order master equation capable of capturing the dynamics of intermediate and strongly coupled systems, which are ubiquitous in microscopic energy transport systems. However, direct implementations of this approach suffer from fundamental scalability issues due to the complexity of the self-consistent equations required to solve for the variational parameters. Here, we present an efficient partitioning scheme that leverages the inherent multi-scale nature of natural energy transport networks. This enables scaling of the variational polaron framework to quantum energy transport systems, constituting hundreds to thousands of sites. Our work unlocks the physically motivated exploration of large transport networks, for example, those present within light-harvesting complexes and exciton transport in disordered semiconductors. 
\end{abstract}
\maketitle

\section{Introduction}
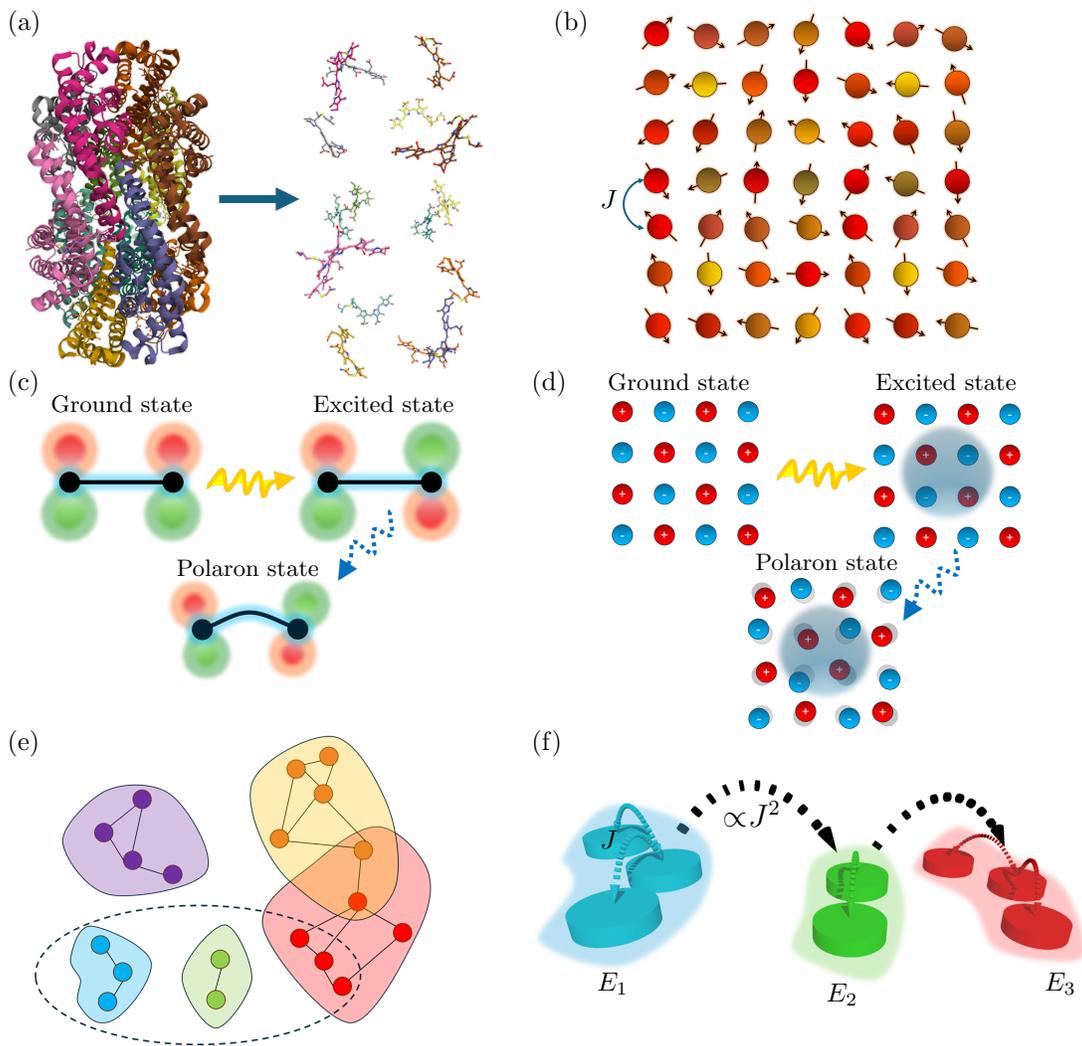
\begin{figure*}[ht]
    \centering
     \begin{overpic}[width=0.81\linewidth]{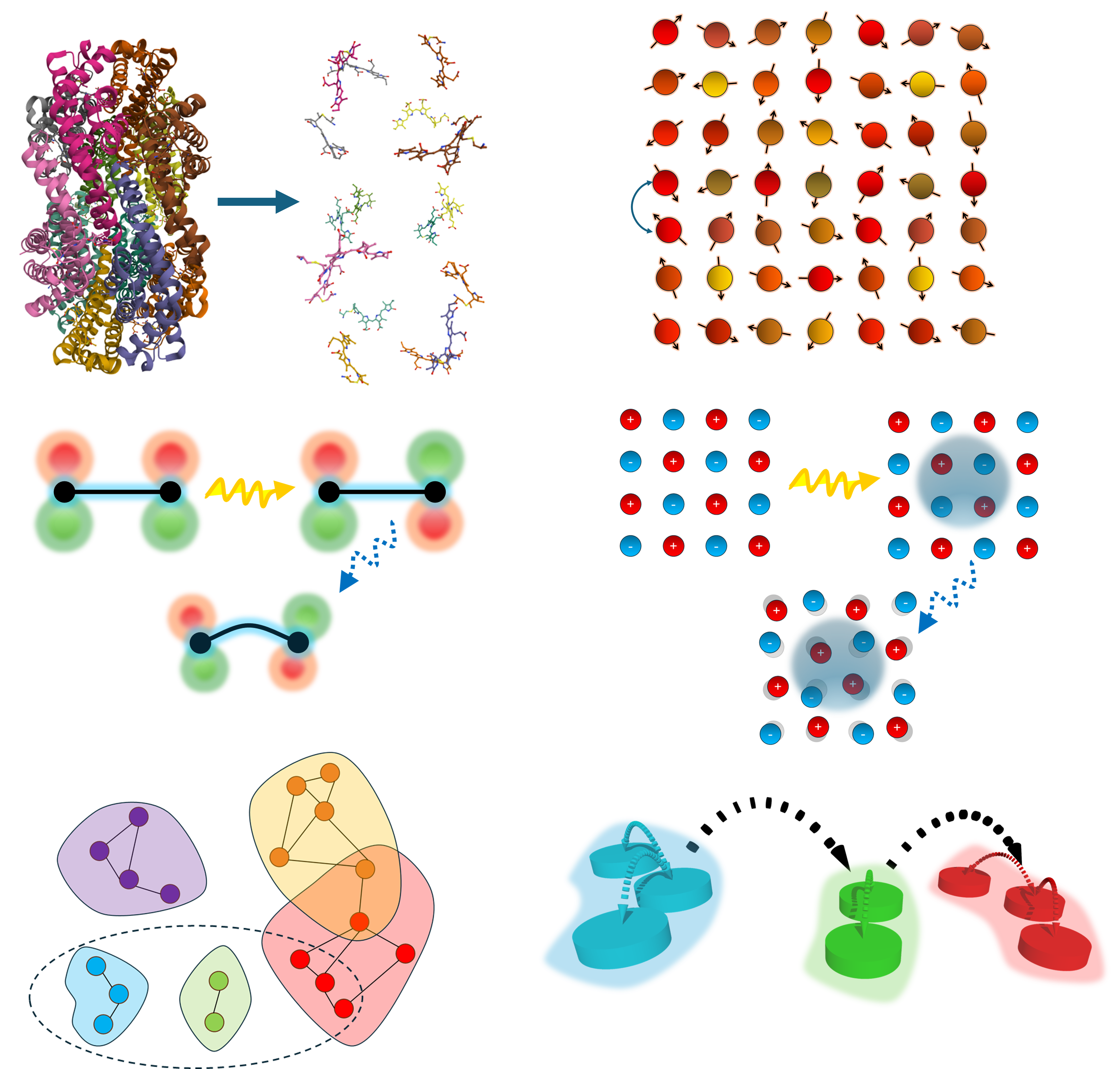}
 \put (0,93) {(a)}
 \put (50,93) {(b)}
 \put (0,60) {(c)}
 \put (4,58) {\small Ground state}
 \put (28,58) {\small Excited state}
 \put (15.5,43) {\small {Polaron} state}
 \put (48,60) {(d)}
 \put (55,60) {\small Ground state}
 \put (79.5,60) {\small Excited state}
 \put (68.5,43.5) {\small {Polaron} state}
 \put (0,27) {(e)}
 \put (48,27) {(f)}
 \put (65.5,20) {\large$\propto\hspace{-3pt} J^2$} 
 \put (55,18) {$\hspace{-3pt} J$} 
 \put (54,5) {$E_1$} 
 \put (75,4) {$E_2$} 
 \put (95,5) {$E_3$} 
 \put (55.,76.7) {$\hspace{-3pt}\small J$} 
 

\end{overpic}
    \caption{(a-b) Two depictions of large quantum energy transfer networks. (a) A portion of the large light harvesting antenna phycobilisomes (PBS) present in cyanobacteria and red algae~\cite{Dodson2022}, showing the removal of the protein scaffold to expose the dipole energy transport network. (b) A disordered material prevalent in organic photovoltaic systems~\cite{Balzer2024}. (c-d) Graphical depictions of polaron formation in (c) molecules (d) atomic lattices. In each case, a photon excites the system from a low energy level to a higher one, though it does not change the nuclear states. After phonon relaxation processes, the composite system equilibrates to the polaron state, where the nuclear degrees of freedom are displaced relative to their ground state configuration. (e-f) Depictions of a quantum exciton transport network with clusters of strongly coupled sites. (e) A schematic of the partitioning of a quantum network into the most relevant blocks for a subset of sites that enables the efficient variational optimisation to be performed. Three of the clusters depicted have site energies $E_1$, $E_2$, and $E_3$ within each cluster and intra-cluster coupling strength $J$. (f) A graphic showing that within these clusters, close molecular subunits will have their dynamics dominated by dipole-dipole interactions, enabling coherent energy transfer to occur (as shown by coloured two-headed arrows depicted with strength $J$). Typically, these have a $r^{-3}$ distance relation. Conversely, for large distance-separated subunits, environmental vibrational effects will dominate, leading to incoherent FRET-type interactions between potentially delocalised subunits (as depicted by single directional black arrows with strength $\propto J^2$ and thus an $r^{-6}$ separation relation). These incoherent processes lead to energy losses, decreasing the system energy and thus following the energy gradient.   }
    \label{fig:Fig1}
\end{figure*}

The dynamics of excitations through nano-scale systems is often greatly influenced by the environment 
leading to decoherence and renormalisation of the excitonic system's energies~\cite{caycedo2022exact}. For example, polaron formation plays a key role in exciton transport in condensed-phase molecular systems, as well as in solid-state materials such as photovoltaic solar cells~\cite{Balzer2023,Balzer2024}. Some archetypal examples of such systems are shown in Fig.\ref{fig:Fig1}(a)-(b). In molecular systems, when a molecule is photoexcited, a transition occurs between electronic states according to the Franck–Condon principle~\cite{Mukamel1995Principles}. As a result, the electronic excitation is initially out of equilibrium with the nuclear degrees of freedom of the molecule. Subsequent vibrational relaxation drives the nuclei toward a new displaced equilibrium configuration, forming a composite quasiparticle consisting of the exciton and the associated nuclear deformation—a ``molecular polaron''~\cite{HOLSTEIN1959325,Munn1985,rouse2019optimal}. A schematic depiction of this process is shown in Fig.~\ref{fig:Fig1}(c).

In solid-state systems, the creation of an exciton within an atomic lattice similarly induces a local lattice distortion, as the surrounding ionic charges are displaced. The exciton, together with this lattice deformation, can likewise be viewed as a quasiparticle polaron~\cite{McCutcheon_2010,nazir2016modelling,Wiercinski2024}, as depicted in Fig.~\ref{fig:Fig1}(d). The physics in both cases manifests itself from nuclei being out of equilibrium after excitation, from which a quasiparticle emerges from the coupling of an exciton to external degrees of freedom that undergo a displacement. Consequently, they can often be treated using related or complementary theoretical approaches.

An important class of naturally occurring transport systems includes photosynthetic complexes such as the Fenna–Matthews–Olsen (FMO) complex in green sulfur bacteria~\cite{Scholes2017,Ishizaki2009,lorenzoni2025microscopicsimulationsuncoverpersistent}, the large light harvesting antenna phycobilisomes (PBS) present in cyanobacteria and red algae~\cite{Dodson2022}, as shown in Fig.~\ref{fig:Fig1}(a), and the chromatophore of purple bacteria comprised of thousands of bacteriachlorophyll molecules~\cite{Sener2007}. Other such systems include the electron transport chain in respiratory complex I~\cite{Martin2017}, olfaction models~\cite{Gane2013}, and bird magnetoreception~\cite{Hamish2016,Gauger2011}.
Beyond quantum effects in biological systems, similar models apply to coupled quantum dots in solid-state physics~\cite{hallett2025controllingcoherencewaveguidecoupledquantum,Wiercinski2024} and impurities in lattice Bose–Einstein condensates~\cite{Klein_2007}.

A plethora of computational techniques have been developed to simulate the dynamics of open quantum systems~\cite{breuer2002theory}.
Some, like path integral methods~\cite{oqupy,Makri2020,SMATPI}, hierarchical equations of motion~\cite{tanimura2020heom}, Dissipation-Assisted Matrix Product Factorization~\cite{Somoza2019}, Thermalised Time Evolving Density matrix using Orthogonal Polynomials Algorithm~\cite {LeDe2024,Prior2010}, Time-evolving matrix product operators~\cite{Strathearn2018,Link2024} and the Automatic Compression of Environments~\cite{Moritz} approaches provide numerically exact results under controlled refinement of convergence parameters.
However, these methods can be computationally demanding, limiting their scope to small-scale simulations, either in system size or environment complexity.
As a result, approximate approaches — such as master equations — are often preferred or indeed the only practical option for the simulation of larger systems~\cite{Weimer2021}.
This alternative approach involves the development of numerically efficient master equations, which describe the propagation of the system after tracing out the environment degrees of freedom. 
While master equations can offer valuable insights by connecting microscopic parameters to rates and energy shifts, they typically rely on perturbative expansions~\cite{REDFIELD19651,breuer2002theory,lindblad_generators_1976,YANG2002355,Shibata1977}, which are contingent on key assumptions about physical parameters that may not hold in particular regimes.
For instance, the ubiquitous Lindblad master equation framework is not capable of capturing the dynamics of non-Markovian or strong system-environment coupling systems~\cite{deVega2017}. 

 

Applying unitary transformations — such as the polaron transformation ~\cite{mahan2013many,Hohenadler2007}, or the reaction coordinate approach~\cite{ReactionCoord} — can suppress the magnitude of the interaction Hamiltonian in a transformed frame, making a perturbative master equation more accurate~\cite{KolliNazir2011,McCutcheon_2010}. This transformation is particularly useful when environmental timescales are short relative to system timescales, or when system-environment coupling is strong while internal system couplings are weak. However, between the weak-coupling and polaron limits, there exists a parameter regime where neither approach is fully effective~\cite{McCutcheon2011_General}. This intermediate coupling regime is typical in molecular energy transfer networks where system geometries and length scales manifest in sub-units coupling to each other with vastly differing coupling strengths~\cite{Ishizaki2009,Engel2007,Ishizaki2012}. A graph of such a network is shown in Fig.~\ref{fig:Fig1}(e), where a subset of the clusters of spatially proximal sites is given. For close molecular subunits, dipole-dipole interactions will be strong, enabling coherent energy transport (an effect captured in weak coupling master equations~\cite{Dubi2020,Davidson2022}). Conversely, for units with greater separation, direct dipole-dipole interactions will be small, and energy transfer will be moderated by incoherent F\"orster Resonant Energy transfer (FRET) processes instead (effects captured in the polaron master equation)~\cite{Forster1959,Jang2008,KolliNazir2011}. However, intermediate distance scales pose challenging conceptual problems where neither weak coupling nor polaron is sufficient, where neither the coherent $r^{-3}$ nor the incoherent $r^{-6}$ relation for transfer rates holds. As such, it is critical to utilise a methodology capable of accounting for weak, intermediate and strong coupling regimes when studying such large molecular networks. A graphical depiction of this concept is shown in Fig.~\ref{fig:Fig1}(f), where we can see coherent short-range effects and incoherent long-range effects. 
The binary choice represented by the two extremes of strong and weak coupling is resolved by the variational polaron transformation~\cite{McCutcheon2011_General,pollock2013multi,Silbey1984,mahan2013many}, which optimises environmental mode displacements based on their frequencies and relative coupling strengths.
This method enables a smooth transition between weak-coupling and polaron descriptions, improving accuracy across parameter regimes. Furthermore, for small systems, the variational master equation has been shown to faithfully generate the dynamics when compared to numerically exact approaches~\cite{McCutcheon2011_General,pollock2013multi}. 

We here present the first fully scalable variational framework that enables simulations of energy-transfer networks at unprecedented scales. This advance is made possible by three key developments: First, we derive a closed-form expression for the variational parameters, eliminating the need for computationally demanding self-consistent solutions. Second, we demonstrate that the global variational optimisation can be systematically reduced to a rapidly convergent set of local optimisations by partitioning the network into its most relevant local substructures, as illustrated in Fig.~\ref{fig:Fig1}(e). Finally, we introduce analytic expressions for the non-Markovian rates appearing in the master equation, dramatically accelerating the evaluation of system dynamics.

Crucially, when combined together, these extensions enable the accurate simulation of networks with hundreds of sites—an achievement previously impossible due to computational limitations of the variational optimisation. This opens new avenues for studying complex quantum transport phenomena in realistic biological and solid-state settings.

This article is organised such that in Section~\ref{sec:VarPol}, we outline the variational polaron transformation. In Section~\ref{sec:ConOpt}, we describe the new convergent local optimisation procedure to greatly enhance the efficiency of the variational optimisation step. Then Section~\ref{sec:Dyn} explores the solution of the variational polaron master equation for specific systems, such as the FMO and LH2 complexes, as well as a model system comprising over one hundred sites, and looks at unique features of these systems enabled by the method developed before concluding in Section~\ref{sec:Conc}. 



\section{Variational Polaron Transformation Model}
\label{sec:VarPol}
The class of systems we shall consider are networks of $N$ (electronic) sites with intersite coupling, typically generated via Coulombic interactions. Additionally, each site in the network has a local environment modelled by a collection of harmonic oscillators. The free electronic Hamiltonian can then be written as 
\begin{equation}
    H_0 = \sum_{n=1}^N E_n\ket{n}\bra{n}+ \sum_{m\neq n}^N V_{nm}\ket{n}\bra{m},
\end{equation}
where $E_n$ is the $n$-th site's excitation energy and $V_{nm}$ is the intersite coupling strength between sites $n$ and $m$. 
The $N$ independent vibrational bath Hamiltonians are 
\begin{equation}
    H_B = \sum_{n=1}^N \sum_\nu \omega_{n\nu} a^\dag_{n\nu}a_{n\nu},
\end{equation}
where $\omega_{n\nu}$ and $a^{(\dag)}_{n\nu}$ are the frequency and annihilation (creation) operator for the $\nu$-mode of the $n$-th site's bath. 

The electronic states couple to their local vibrational environments via a state-dependent linear coupling of the form
\begin{equation}
    H_{I} = \sum_{n=1}^N \sum_\nu g_{n\nu}\ket{n}\bra{n}(a^\dag_{n\nu}+a_{n\nu}),
\end{equation}
where $g_{n\nu}$ characterises the coupling strength of the $n$-th site to the $\nu$-mode in its environment. 
Each vibrational environment has its frequency dependent coupling determined by the spectral density  
\begin{equation}
    J_n(\omega) = \sum_\nu g_{n\nu}^2\delta(\omega-\omega_{n\nu}).
\end{equation}

As discussed above, many techniques have been developed to study such systems, ranging from perturbative, limited in their applicability, to numerically exact frameworks that are often computationally expensive. Each, however, has its own regimes of applicability. 
To combat the limitations of the perturbative frameworks, we utilise a variationally optimised polaron transformation to develop a master equation that has the complexity of perturbative approaches while effectively interpolating between weak, intermediate and strong coupling regimes. As such, we deploy a partial polaron transformation 
\begin{eqnarray}
\tilde{H} = U_PHU_P^\dag = e^G H e^{-G},  \label{eq:pol}
\end{eqnarray}
where
\begin{eqnarray}
    G = \sum_{n\nu}{\ketbra{n}{n} \frac{f_{n\nu}}{\omega_{n\nu}} ( b^\dagger_{n\nu} -  b_{n\nu})},
\end{eqnarray}
and where $f_{n\nu}$ are free parameters determining by how much each mode is displaced.
This transforms our system into a displaced frame of the vibrational environments, which depends on the state of the system. For $f_{n\nu}=g_{n\nu}$, we obtain the full polaron transformation that diagonalises the interaction Hamiltonian, and for $f_{n\nu}=0$ we have the 
undisplaced weak coupling frame. Performing the unitary transformation on our Hamiltonians generates the following transformed frame Hamiltonians: the variational frame electronic Hamiltonian
\begin{alignat}{3}
    \tilde{H}_{0}	&= &&\tilde{E} \hspace{33pt} +  &&\tilde{V} \nonumber\\ &=
    \sum_{n} &&\tilde{E}_n \ketbra{n}{n} + \sum_{n \neq m} &&\tilde{V}_{nm} \ketbra{n}{m},
\end{alignat}
where $\tilde{V}_{nm}=\mathcal{B}_{n}\mathcal{B}_{m} V_{nm}$ are the renormalised intersite couplings and $\tilde{E}_n = {E}_n + R_n$ are the renormalised site energies. 
The displacement of the vibrational modes introduces a reorganisation of the site energies through $R_n$, and the partial formation of polarons suppresses the intersite couplings through $0\leq\mathcal{B}_n\leq 1$, leading to greater localisation of eigenstates. These two renormalisations of the energies take the form 
\begin{align}
\mathcal{B}_n &= \tr\left\{ B_n \rho_B \right\} \nonumber \\
&= \exp\left[-\frac{1}{2} \sum_{\nu}{\frac{f_{n,\nu}^2}{\omega_{n,\nu}^{2}} \coth(\beta \omega_{n,\nu}/2)}\right]\label{eqn:Bn},
\end{align}
where $\rho_B=\exp\{-\beta H_B\}/\tr{\exp\{-\beta H_B\}}$ is the bath Gibbs state at inverse temperature $\beta=1/k_BT$, and where the individual site bath displacement operators are defined as 
\begin{eqnarray}
 B_{n} = \exp\left[\sum_{\nu}\frac{f_{n,\nu}}{\omega_{n,\nu}}{ ( b^\dagger_{n,\nu} -  b_{n,\nu})}\right]  ,
 \end{eqnarray}
and the site reorganisation energies are defined by 
\begin{eqnarray}
R_n = \sum_{\nu}\frac{f_{n,\nu}}{\omega_{n,\nu}}{\left[f_{n,\nu}-2 g_{n,\nu}\right]}. \label{eq:rpars}
\end{eqnarray}

The interaction Hamiltonian now also obtains additional terms, some from the original coupling and some from the intersite couplings
\begin{align}
 \tilde{H}_{I} &= \tilde{H}_{L} + \tilde{H}_{D},   \\
\tilde{H}_{L}	&= \sum_{n\nu}{\ketbra{n}{n} (g_{n\nu}-f_{n\nu})  \left[a^\dagger_{n\nu} + a_{n\nu}\right]},  \\
\tilde{H}_{D}&= \sum_{n \neq m}{ V_{nm} \ketbra{n}{m} B_{nm}}, \\\tilde{H}_B &= H_B .
\end{align}
The terms $\Tilde{H}_L$ are the remnant terms from the partial diagonalisation of the linear coupling to the vibrational environment, while the $\Tilde{H}_D$ are responsible for phonon-assisted hoppings between the sites, generating FRET-like dynamics. These off-diagonal bath coupling operators are defined by the bath displacement operators for the two coupled sites
\begin{eqnarray}
 B_{nm} = B_{mn}^\dagger = B_nB_m^\dag - \mathcal{B}_{n}\mathcal{B}_{m} .  
\end{eqnarray}
We treat the displacements as variational parameters, determining them through a free-energy minimisation procedure following the approach originally introduced by Silbey and Harris~\cite{Silbey1984}. The free energy of the system and environment can be expressed in the following way using a Taylor expansion in $\tilde{H}_I$
\begin{equation}
A = -\frac{1}{\beta} \ln \left[ \tr{e^{-\beta \tilde{H}_F}} \right] + \langle \tilde{H}_I \rangle_{\tilde{H}_F} + \langle\mathcal{O}\left( \tilde{H}_I^2  \right)\rangle_{\tilde{H}_F},
\label{eq:FBbound}
\end{equation}
where $\tilde{H}_F = \tilde{H}_0+\tilde{H}_B$ is the combined free system and bath Hamiltonians. 
By construction of our interaction Hamiltonian $\tilde{H}_{I}$, the first-order term is zero, and as such, we can approximate the free energy with the bound on the free energy
\begin{equation}
    A\approx A_B = -\frac{1}{\beta}( \ln \left[ \tr{e^{-\beta \tilde{H}_0}} \right] + \ln \left[\tr{ e^{-\beta \tilde{H}_B}}\right]).
\end{equation}

This leads to the variational optimisation 
\begin{equation}
\frac{\partial A_B}{\partial f_{n\nu}}  = 0 \qquad \forall \; n, \nu.
\end{equation}
Since $\tilde{H}_B$ has no dependence on $f_{n,\nu}$, it follows that we need only perform the optimisation on the system free energy term.
Performing such an optimisation of the network leads to the calculation of the set of variational parameters $\{ f_{n\nu}\}$. By relating the displacement with the coupling strength of the vibrational modes $f_{n,k}=F_n(\omega_{n,k})g_{n,k}$, where we have introduced the variational displacement function $F_n(\omega)$, the variational optimisation determines the form of this function to be
\begin{equation}
    F_n(\omega) = \frac{\omega}{\omega+\alpha_n \coth(\beta\omega/2)}.
\end{equation}
This defines a function $0\leq F_n \leq 1$ where the variational parameter $\alpha_n$ must be solved for each site. For large values of $\omega\gg \alpha_n$, $F_n(\omega)\rightarrow 1$, and for small $\omega\ll\alpha_n$ $F_n(\omega)\rightarrow 0$. This shows that low-frequency modes are treated via an untransformed weak-coupling type master equation, whereas a polaron-type master equation is used to handle higher-frequency modes. The variational displacement function also has low-frequency behaviour $F_n(\omega)\sim \beta\omega^2/2\alpha_n$, removing the divergences present for Ohmic spectral densities suffered by conventional polaron theory, such as in Eqn.~\ref{eqn:Bn} (see Appendix~\ref{sec:master} for more detail). Further, when $\alpha_n$ is large compared to vibrational frequencies, the variational displacement function will be small, leading to an undisplaced frame and effectively weak coupling dynamics. Conversely, when $\alpha_n$ is small compared to environmental frequencies, $F_n$ will be close to one, and the frame will be mostly displaced, resembling a standard polaron transformed frame. By utilising properties of the derivatives of matrices and the cyclic property of the trace, we calculate, for the first time, a succinct form for the variational parameter 
\begin{equation}
    \alpha_n = -\frac{\tr\{\ketbra{n}{n}\Tilde{V} \tilde{\rho}_S^\beta\}}{\tr\{\ketbra{n}{n}\tilde{\rho}_S^\beta \}},\label{eqn:alpha}
\end{equation}
where $\rho_S^\beta$ is the free Gibbs state of the system $\rho^\beta_S = e^{-\beta \Tilde{H}_0}/\tr{e^{-\beta \Tilde{H}_0}}$. This compact form of the variational parameter substantially reduces the numerical complexity of the optimisation procedure by eliminating the need for computationally expensive numerical differentiation.
This form has an intuitive interpretation as being the delocalisation caused by the system Hamiltonian of the $n$-th site, divided by the probability of being in the $n$-th site at the thermalised state $\tilde{\rho}_S^\beta$. For large amounts of delocalisation (large $V_{nm}$), $\alpha_n$ will be large, resulting in a mostly unperturbed frame, where we anticipate weak coupling to be appropriate. Conversely, for largely localised states, where $V_{nm}$ is small, $\alpha_n$ will be small too, leading to a mostly displaced and polaron-like frame, as expected. 
$\alpha_n$ must then be calculated via the set of self-consistent equations generated by Eqns.~\ref{eqn:Bn}, \ref{eq:rpars} and \ref{eqn:alpha}, and once calculated, can be used to generate the time-convolutionless second-order perturbative variational polaron master equation shown in Appendix~\ref{sec:master}.

This approach generates a master equation that is capable of interpolating between coupling strength regimes~\cite{McCutcheon2011_General,pollock2013multi}. 
However, determining the variational parameters becomes more challenging as the network gets larger. This is due to repeated calculation of the total system free energy, requiring many large matrix exponentials to be performed and the optimisation to be undertaken for each site in the network. Often, this can be computationally more expensive than calculating the resulting dynamics, even for small systems. In the following section, we outline another advance through an efficient convergent approach for calculating the variational parameters.

\section{Convergent Local Optimisation Within Networks}
\label{sec:ConOpt}
To ameliorate the poor scaling of the full variational optimisation for increasing network sizes, we propose a site-local optimisation. We start by considering a partitioned version of the free energy that approximates the free energy about an individual site. This is achieved by considering different partitions around the particular site we wish to optimise of size $p$, such that when we calculate the free energy, we need only include the $p-1$ most relevant other sites. To prove the efficacy of this approach, we define the site-local free Hamiltonian as
\begin{align}
    H^{L}_{i,p} &= \mathcal{P}_{i,p} \tilde{H}_0= \tr{\tilde{H}_0}_{\beta_{i,p}}\hspace{-1pt}\otimes \mathbf{1}^{N-p}\nonumber\\
    &=\hspace{26pt}\tilde{ E}^L_{i,p} \hspace{21pt}+ \hspace{39pt}\tilde{V}_{i,p}^L
    \nonumber\\
     &= \sum_{n\in \alpha_{i,p}}{\tilde{E}_n \ketbra{n}{n}} + \sum_{n \neq m\in \alpha_{i,p}}{ \tilde{V}_{nm} \ketbra{n}{m}}, 
\end{align}
where $\mathcal{P}_{i,p}$ is a projection operator onto the subspace of sites in the set $\alpha_{i,p}$, the set of so-called `relevant' site indices to the site at index $i$, given by tracing over the complement set of `irrelevant' sites $\beta_{i,p}
\equiv \{1,..,N\}\setminus\alpha_{i,p} $.  As we shall see, it is appropriate to determine these relevant sites by finding the $p-1$ other sites most strongly coupled to the $i$-th site, i.e., those with the largest $|V_{ij}|$.
Conversely, we can consider the `irrelevant' Hamiltonian as 
\begin{align}
    H^{\text{IR}}_{i,p} &= \mathcal{I}_{i,p} \tilde{H}_0 = \mathbf{1}^{p}\otimes  \tr{\tilde{H}_0}_{\alpha_{i,p} }\nonumber\\
    &=\sum_{n\in \beta_{i,p}}{\tilde{E}_n \ketbra{n}{n}} + \sum_{n \neq m\in \beta_{i,p}}{ \tilde{V}_{nm} \ketbra{n}{m}}, 
\end{align}
where we have instead traced the system Hamiltonian over the `relevant' sites.
\begin{figure}[t]
    \centering
         \begin{overpic}[width=\linewidth]{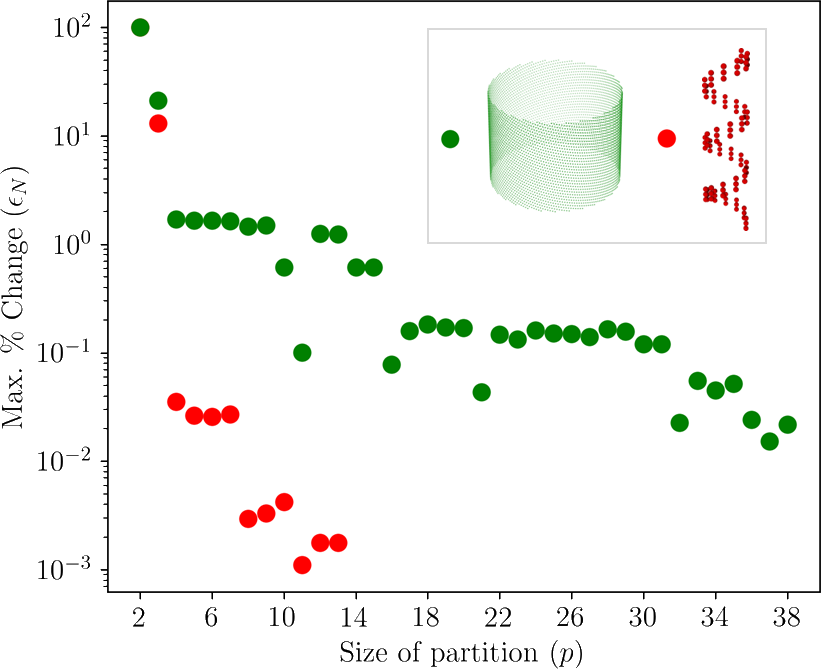}
 
 
\end{overpic}    
    \caption{A plot to show the convergence of the variational parameters for increasing partition size. $\epsilon_N$ is the maximum relative difference across all variational parameters $\{R_n,\mathcal{B}_n\}$ between partition size $N-1$ and $N$. The two systems considered are a model of the rod-like chlorosome of green sulfur bacteria (green), comprising 5400 bacteriochlorophyll dipoles and a 3000 site model energy transport network comprised of triplets in a helical structure with detuning along its length (red). A depiction of each system is shown in the inset.  See Appendix~\ref{app:Params} for more details on the Hamiltonians. This plot shows convergence for relatively small partition sizes compared to the total network size. This convergence behaviour is not necessarily monotonic with the partition size: relative changes in the variational parameters are observed to cluster, notably into triplets for the helical model. This directly reflects the underlying triplet structure of the network, demonstrating that the effectiveness of the partitioning scheme is intrinsically dependent on the network geometry.}
    \label{fig:variationalError}
\end{figure}

From these projected Hamiltonians, we can define the `connection' Hamiltonian that couples these two partitions of our Hilbert space as
\begin{align}
    H^{C}_{i,p} &= \Tilde{H}_0 -H^{\text{IR}}_{i,p} -H^{L}_{i,p} \nonumber\\
    &= \sum_{n \in \alpha_{i,p} ,m\in \beta_{i,p}}{ \tilde{V}_{nm} (\ketbra{n}{m}} + \ketbra{m}{n}).\label{eqn:connection}
\end{align}
It is trivial to see that the relevant and irrelevant Hamiltonians commute with each other as they act on disjoint Hilbert spaces, but neither commute with the connection $H^{C}_{i,p}$; this property will be useful later.
We return now to the definition of the free energy bound for the system
\begin{align}
    A_B &= -\frac{1}{\beta} \ln \left[ \tr(e^{-\beta \tilde{H}_0}) \right] \\&= -\frac{1}{\beta} \ln \left[ \tr(\exp{-\beta( H^{\text{IR}}_{i,p} +H^{L}_{i,p}+H^{C}_{i,p})} )\right].\nonumber
\end{align}
By a suitable choice of our partitioning for a site $i$, we can assume that the connection Hamiltonian is relatively small, such that we can perform a Taylor expansion in $H_{i,p}^{C}$ and thus the free energy is given by

\begin{gather}
    A_B = -\frac{1}{\beta} \ln \left[ \mathcal{Z}_A+\mathcal{E} \right]. 
\end{gather}
The zeroth-order term of the partition function is given by a product of our relevant and irrelevant degrees of freedom partition functions
\begin{gather}
  \mathcal{Z}_A=\tr{e^{-\beta H^{\text{IR}}_{i,p}}}\tr{e^{-\beta H^{L}_{i,p}}},
\end{gather}
and all higher order terms in $H_{i,p}^{C}$ are encapsulated in $\mathcal{E}$. For a small connection Hamiltonian, we make a linear approximation of this error, which has the form
\begin{equation}
    \mathcal{E} = -\beta\tr{\exp{-\beta H^{\text{IR}}_{i,p}}\exp{-\beta H^{L}_{i,p}}H^{C}_{i,p}}.
\end{equation}
Further Taylor expansion of the logarithm to first order in the error $\mathcal{E}$, yields
\begin{align}
    A_B = &-\frac{1}{\beta} \left(\ln\left[\tr{e^{-\beta H^{\text{IR}}_{i,p}}}\right]+\ln\left[\tr{e^{-\beta H^{L}_{i,p}}}\right]\right) \nonumber\\&- \frac{\mathcal{E}}{\beta\mathcal{Z}_A}. 
\end{align}
Performing an expansion in the Hamiltonians, we find that the lowest order contribution has the form 
\begin{equation}
    \frac{\mathcal{E}}{\beta \mathcal{Z}_A} = \frac{\beta}{\mathcal{Z}_A} \sum_{n\in \beta_{i,p},m\in\alpha_{i,p}} |\tilde{V}_{nm}|^2= \frac{\beta}{\mathcal{Z}_A}||H^C_{i,p}||^2_F,
\end{equation}
where $||\cdot||_F$ denotes the Frobenius norm. This clearly shows that this is, in fact, a correction up to second order in the connection Hamiltonian. 

By construction, we anticipate that the coupling of sites within the partition to those outside the partition is small, and thus, this error is well controlled. 
This justifies our neglecting the linear error from our optimisation. 
Further, by construction, the irrelevant Hamiltonian has no dependence on $f_{i,\mathbf{k}}$, and as such, the minimisation
\begin{gather}
    \frac{\partial A_B}{\partial f_{i\nu}} = 0
    \implies \frac{\partial}{\partial f_{i\nu}} \ln\left[\tr{e^{-\beta H^{L}_{i,p}}}\right]= 0, 
\end{gather}
is satisfied. 
This suggests that for a necessarily small connection Hamiltonian, it is sufficient to minimise the free energy of the relevant partition for each site and not that of the entire network. Using this approximation, we can modify our self-consistent equations such that 
\begin{equation}
    \alpha_n = -\frac{\tr\{\ketbra{n}{n}\tilde{V}_{n,p}^L \rho_{nLS}^\beta\}}{\tr\{\ketbra{n}{n}\rho_{nLS}^\beta \}},\label{eqn:alphaLocal}
\end{equation}
where $\rho_{nLS}^\beta$ is now the free Gibbs state with respect to the local Hamiltonian ${H}_{n,p}^L$ such that $\rho_{nLS}^\beta = e^{-\beta H^L_{n,p}}/\tr{e^{-\beta H^L_{n,p}}}$. This means that the matrix exponentials and multiplications need only be performed on matrices of dimensionality $p\times p$ instead of the full electronic Hilbert space Hamiltonians with dimension $N\times N$. These need to be performed $N$ times (once per site) for each step of the self-consistent equation solver; as such, we anticipate that a naive implementation would have a complexity of $\mathcal{O}(N p^3)$ for the local optimisation vs $\mathcal{O}(N^4)$ for the original global optimisation. As we will show later, in many cases we can attain convergence with $p\ll N$, making this implementation considerably more advantageous.  
The efficacy of this approach is predicated on choosing appropriate partitions of size $p$ such that the connection Hamiltonian is minimised. We can also see from the definition of the connection Hamiltonian in  Eqn.~\ref{eqn:connection} that it is natural to determine the most relevant sites based on the magnitude of inter-site coupling $|V_{nm}|$ such that those that remain in the connection Hamiltonian are negligibly small. In Fig.~\ref{fig:Fig1}(c), we show a schematic of how a subset of these partitions of a quantum network may look; the coloured regions represent a single site's most relevant sites partition. Note that we need not for these partitions to be disjoint or symmetric, just that they sufficiently minimise that particular site's connection Hamiltonian; however, if these do hold, the complexity of the solution reduces further. 
Utilising this method, we calculate the variational optimisation of two quantum energy transport networks. The first is a model of the rod-like chlorosomes inside green sulfur bacteria; such a network is composed of 5400 sites and exhibits highly directional energy transport. The second system is a helical dipole system network with 3000 sites to validate this new procedure. Such a procedure would be entirely intractable with the previously devised full system free energy optimisation. We determine convergence by increasing the size of the partitions $p$ and stopping when increasing the partition size has minimal effect (e.g. no changes above 1 part in 2000). The results of these optimisations are shown in Fig.~\ref{fig:variationalError}, where we can see that for the 3000-site spiral toy model system, the maximum relative difference for all $\mathcal{B}_n$ and $R_n$ as we increase the partition size for the variational optimisation quickly converges, to maximum relative errors of less than $10^{-3}$ for fewer than 5 sites. We see similar behaviour for the rod-like structure of the green sulfur bacteria, though convergence is slower, due to the greater number of nearby dipoles. However, the required number of sites in the partition is vastly fewer than the total number of sites in both networks. We also note that the structure of the relative errors is grouped into triplets for the helical network; this is due to the geometric structure of the toy-model system, comprising strongly coupled trimers. These results showcase that appropriate clustering of sites within partitions enables efficient calculation of variational parameters, demonstrating the power of this procedure. The parameters for this network are outlined in Appendix~\ref{app:Params}. 

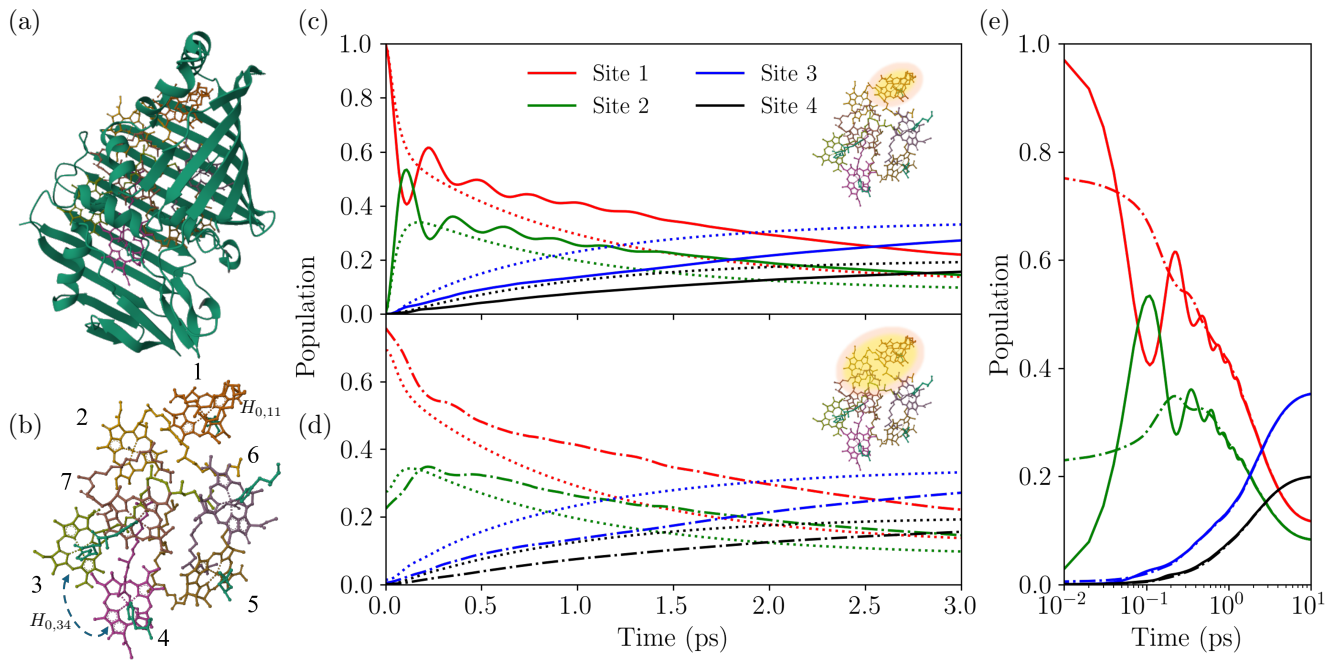
\begin{figure*}[t]
    \centering
     
    \begin{overpic}[width=\linewidth]{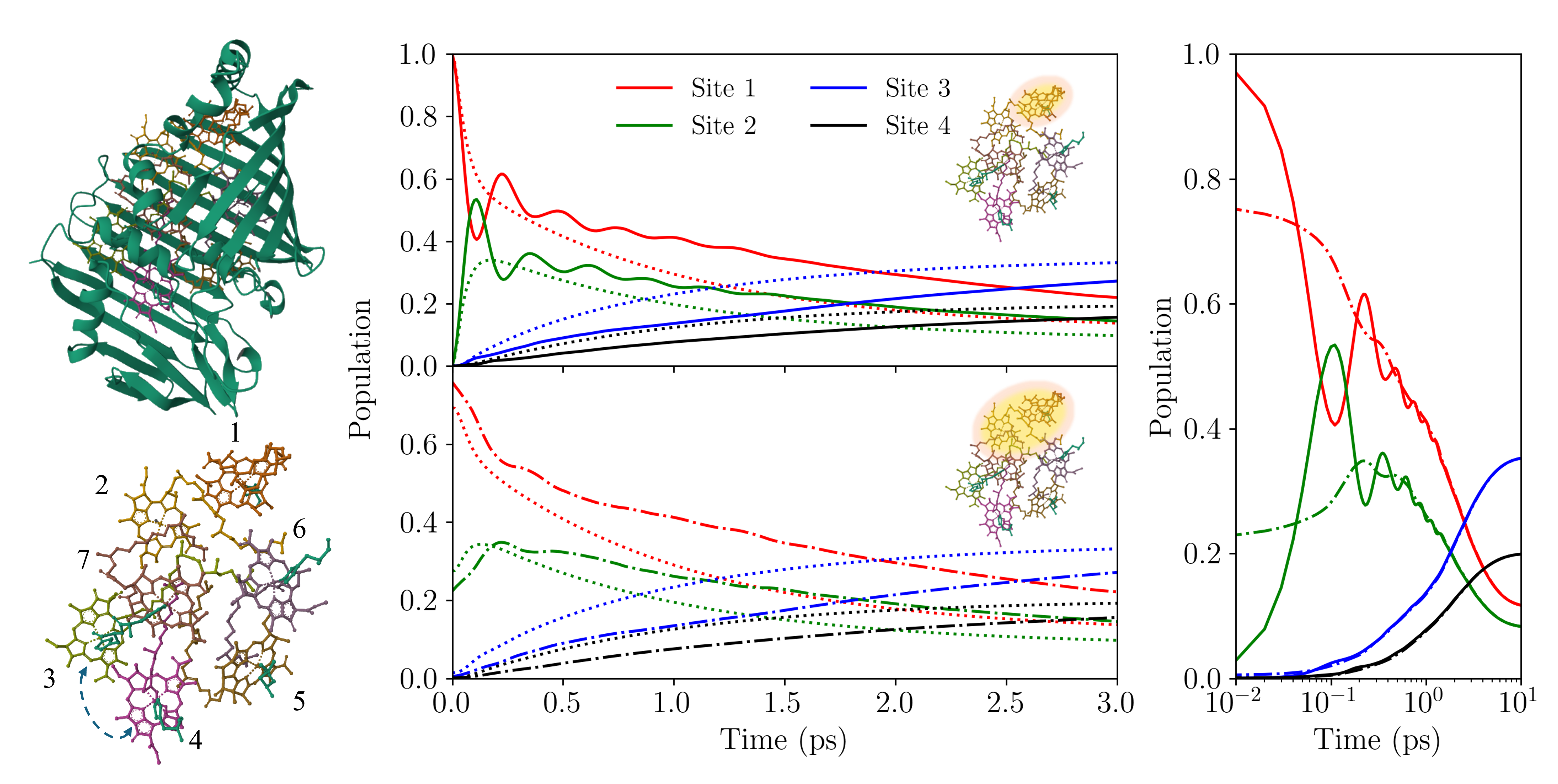}
     \put (1,48) {(a)}
     \put (22.5,48) {(c)}
     \put (18.2,19.5) {\scalebox{0.7}{$H_{0,11}$}}
     \put (2.6,3.7) {\scalebox{0.7}{$H_{0,34}$}}
     
     \put (22.5,18) {(d)}
     \put (73,48) {(e)}
     \put (1,18) {(b)}
    \end{overpic}
    \caption{(a) Crystal structure of the monomeric FMO complex. (b) Removing the protein scaffold yields the seven BChl a pigments in the FMO monomer, labeled by sites 1 to 7. These sites comprise the energy network generating the system Hamiltonian determined by the intersite couplings for the off-diagonal terms  ($H_{0,ij}$) and the individual site energies for the diagonal components ($H_{0,ii}$). (c),(d) Dynamics of the FMO site populations using the non-Markovian variational frame master equation (thick (c) dashed dot (d)) and weak coupling Bloch-Redfield master equation (dotted), with (c) the first site initially excited (d) the energy eigenstate with greatest support on the first site initially excited. (e) The dynamics of the variational polaron master equation populations from (c) and (d) are superimposed, with the localised initial state (thick) and the delocalised variational polaron eigenstate initial condition (dashed dotted), showing early time deviations and convergence after a few picoseconds.}
    \label{fig:FMODyn}
\end{figure*}
\section{Non-Markovian intermediate scale dynamics of large networks}
\label{sec:Dyn}
Having developed a framework for computing the variational parameters required in the variational polaron master equation, we now apply it to calculate the dynamics of large energy-transfer networks. The variational master equation can be calculated using the second-order time convolutionless master equation in the variational polaron frame
\begin{equation}
\frac{d}{dt}\Tilde{\rho}_S = -i[\Tilde{H}_0,\rho_S] -\tr_B\{\int^t_0 ds[\tilde{H}_I(t),[\tilde{H}_I(s),\Tilde{\rho}_S\otimes\rho_B]]\}, 
\end{equation}
where $\tilde{\rho}_S$ is the reduced density matrix for the system in the variational polaron frame. The full derivation of this was performed in~\cite{pollock2013multi}, and for completeness, we reproduce it in Appendix~\ref{sec:master}. Furthermore, in Appendix~\ref{app:Rates}, we have calculated analytic solutions for the non-Markovian rates that define the master equation, as typically these are numerically challenging. Utilising this master equation framework, we proceed to investigate several test systems of increasing complexity. To demonstrate the efficacy of our approach and compare the resulting dynamics with those obtained from a conventional weak-coupling Bloch-Redfield master equation without Lamb shifts~\cite{breuer2002theory}. Furthermore, the computational simplicity of the master equation formalism allows us to probe novel features of these systems, including the dynamics dependence on initial conditions, the influence of individual vibrational modes on energy transport, and the emergence of transitions from delocalised to localised eigenstructures.

\begin{figure*}[t]
    \centering
     
    \begin{overpic}[width=0.95\linewidth]{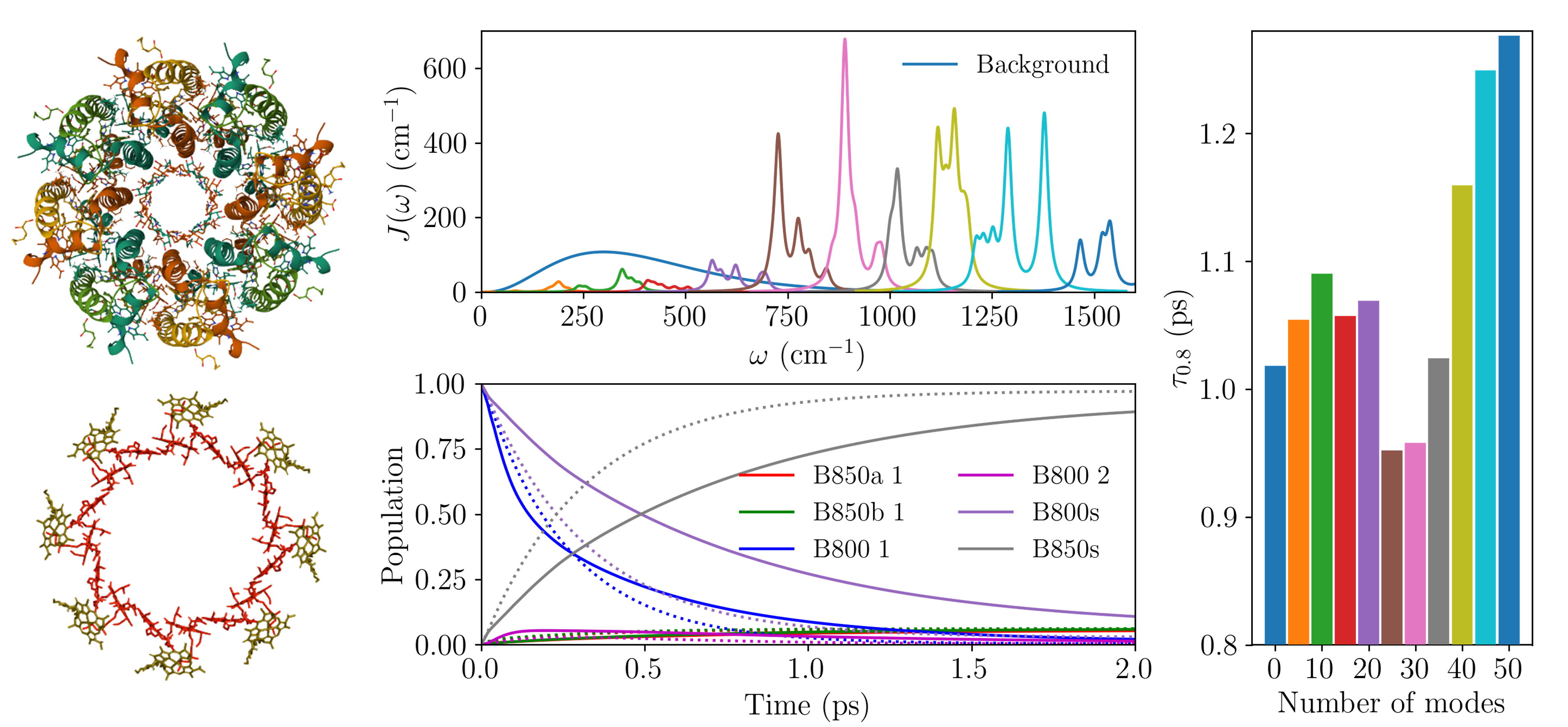}
     \put (0,44) {(a)}
     \put (23,44) {(c)}
     \put (23,21) {(d)}
     \put (77,44) {(e)}
     \put (0,21) {(b)}
     \put (8.5,15) {\small B850s}
     \put (8.5,1.5) {\small B800s}
     \put (3,5.5) {\small 1}
     \put (4,10) {\small a}
     \put (7.5,9.5) {\small b}
     \put (0.5,13) {\small 2}
    \end{overpic}
    \caption{(a) Crystal structure of the LH2 complex from Rhodospirillum molischianum. (b) Removing the protein scaffold yields the twenty-four BChl pigments in the outer ring of higher transition energy 800 nm species (yellow) and an inner ring of lower 850 nm species (red). (c)  The spectral density used for studying the dynamics of the LH2, comprising fifty normal modes and a broad background, we have grouped the normal modes into groups of five and show the broad background spectrum. (d) Dynamics of excitation in a LH2 complex of Rhodospirillum molischianum solved using weak coupling Bloch-Redfield (dashed) and variational polaron master equation (thick) with an outer B800 molecule initially excited. (e) A bar graph depicting the time taken to reach $80\%$ population in the inner ring B850 molecules, starting with an excitation on the outer ring for adding normal modes from the environment, the colours of the plots depict which normal modes have been added.}
    \label{fig:LH2Dyn}
\end{figure*}
\subsection{FMO Complex}
The first system we consider is the seven-site monomer of the Fenna-Matthews-Olson (FMO) complex, shown in Fig.~\ref{fig:FMODyn}(a) with its protein scaffold and in Fig.~\ref{fig:FMODyn}(b), where only the bare bacteriochlorophyll (BChl) pigments are depicted along with their site indices. The FMO complex is a canonical model system in the study of quantum effects in biological processes~\cite{Engel2007,lorenzoni2025microscopicsimulationsuncoverpersistent}, due to the appearance of long-lived coherences in 2D spectroscopy experiments. Whether or not such coherences play a functional role remains a subject of debate~\cite{Cao2020}. 

For the dynamics studied here, the untransformed system Hamiltonian has values~\cite{FMO_Ham}
\begin{widetext}
\begin{equation}
H_{0} = 
\begin{pmatrix}
240 & -87.7 & 5.5 & -5.9 & 6.7 & -13.7 & -9.9 \\
-87.7 & 315 & 30.8 & 8.2 & 0.7 & 11.8 & 4.3 \\
5.5 & 30.8 & 0 & -53.5 & -2.2 & -9.6 & 6.0 \\
-5.9 & 8.2 & -53.5 & 130 & -70.7 & -17.0 & -63.3 \\
6.7 & 0.7 & -2.2 & -70.7 & 285 & 81.1 & -1.3\\
-13.7 & 11.8 & -9.6 & -17.0 & 81.1 & 435 & 39.7\\
-9.9 & 4.3 & 6.0 & -63.3 & -1.3 & 39.7 & 245
\end{pmatrix} 
{\rm cm}^{-1}.
\end{equation}
\end{widetext}
As is apparent from this Hamiltonian, site 3 has the lowest energy; we thus expect this site to have the greatest population in the thermalised state. 
The vibrational environment of FMO is highly structured and, using normal-mode decompositions, 62 modes can be determined to be the most significant to the system's evolution~\cite{lorenzoni2025microscopicsimulationsuncoverpersistent,Duan2017}. These normal modes and their associated Huang-Rhys factors, $S_n$, were used to define the spectral density 
\begin{equation}\label{eq:specnormal}
    J_\text{N}(\omega)=\sum_{n=1}^{62} S_n \frac{4\omega\gamma\omega_n(\omega_n^2+\gamma^2)}{\pi((\omega-\omega_n)^2+\gamma^2)((\omega+\omega_n)^2+\gamma^2)},
\end{equation}
with $\gamma=5~\text{cm}^{-1}$. We also added a broad background Adolphs-Renger spectral density to account for additional low-frequency environmental noise 
\begin{eqnarray}
    J_{\text{B}}(\omega) =\frac{S}{s_1+s_2}\sum_{i=1}^2\frac{s_i}{7!2\omega_i^4}\omega^5\exp\{-(\omega/\omega_i)^{1/2}\},
\end{eqnarray}
with $S = 0.29$, $s_1=0.8$, $s_2=0.5$, $\omega_1=0.056$ cm$^{-1}$  and $\omega_2=1.94$ cm$^{-1}$~\cite{RATSEP2007251,lorenzoni2025microscopicsimulationsuncoverpersistent}.
Current numerically exact methods struggle to simulate the full seven-site system with all 62 normal modes~\cite{lorenzoni2025microscopicsimulationsuncoverpersistent} due to the complexity and size of the environment. By contrast, within the variational polaron framework, the corresponding master equation can be solved within seconds on a standard personal computer.

The results of these simulations are shown in Fig.~\ref{fig:FMODyn}(c), where site~1 is initially excited. We observe transient oscillatory dynamics at early times, followed by a transition to exponential-like decay. This oscillatory behaviour is absent in the weak-coupling Bloch-Redfield results shown alongside, indicating that the decoherence is overestimated in the weak coupling case, leading to damping of these oscillations. Moreover, the weak-coupling theory substantially overestimates the decay rates in the system.

A key advantage of the variational approach is its computational efficiency, allowing exploration of many initial conditions and environmental configurations. In optical experiments, laser pulses typically excite energy eigenstates rather than individual sites~\cite{Mukamel1995Principles}. In Fig.~\ref{fig:FMODyn}(d), we therefore show the dynamics when the initially excited state is the variational polaron eigenstate with the largest localisation on site~1. Since this is an eigenstate of the system, it does not exhibit coherent oscillations arising from the system Hamiltonian; thus, any oscillations in the population dynamics arise solely from non-Markovian vibrational effects. We note that the dynamics for the delocalised eigenstate and those for the localised state appear similar after 2 ps. As such, Fig.~\ref{fig:FMODyn}(e) compares the population dynamics for the two initial conditions considered. Despite one being a fully localised site excitation and the other a delocalised eigenstate, their dynamics converge on a timescale of approximately 2~ps. All simulations were performed with a phonon environment at 300\,K.

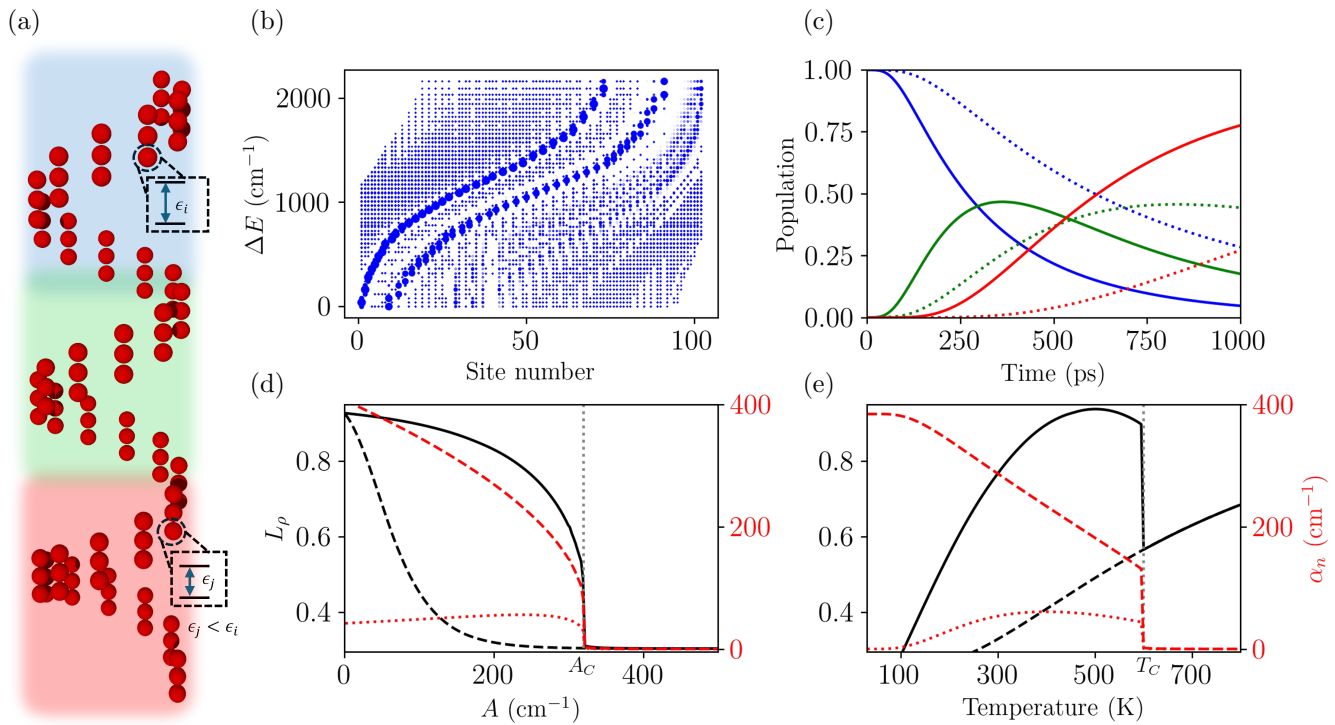
\begin{figure*}[t]
    \centering
    \begin{overpic}[width=\linewidth]{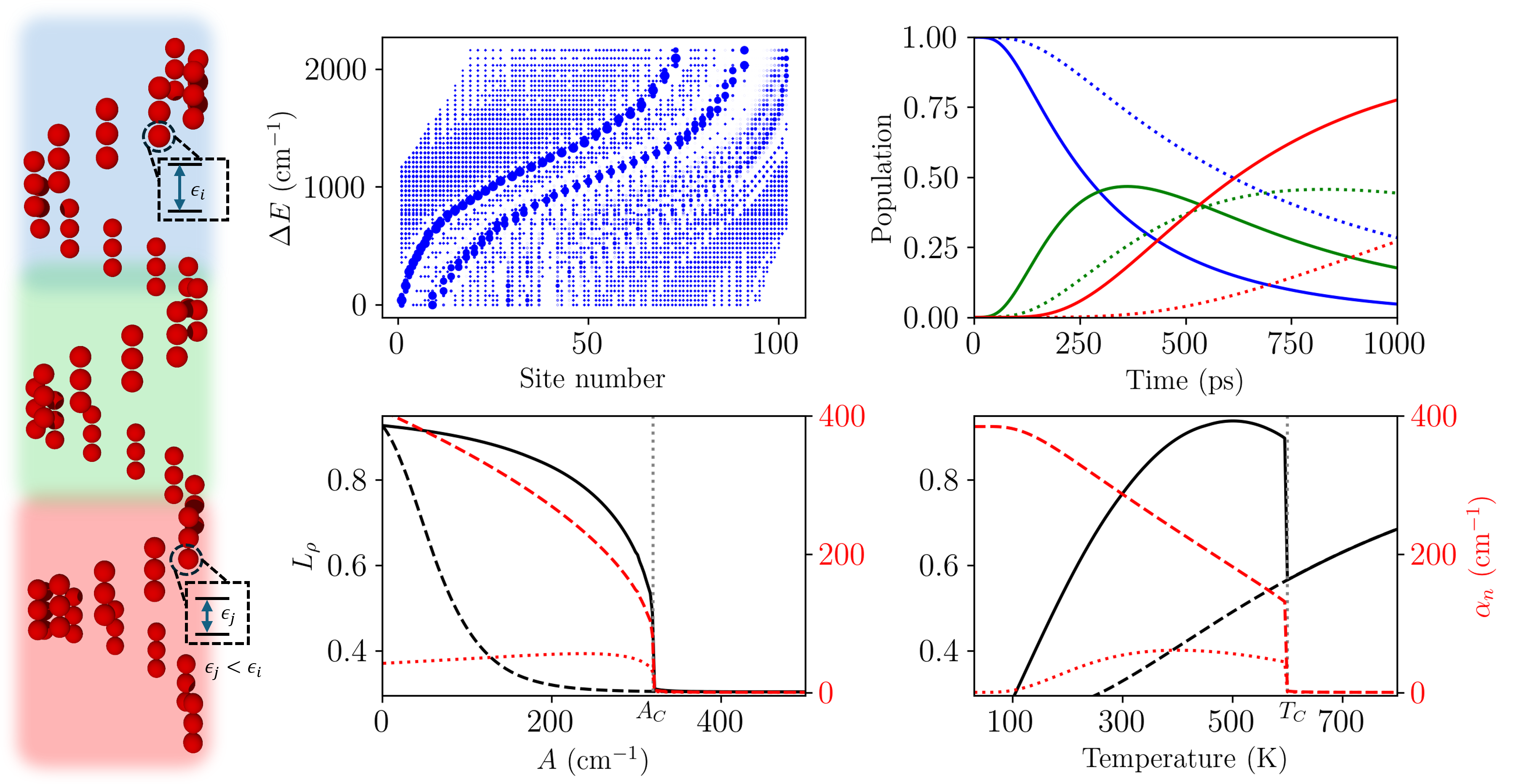}
     \put (0,52) {(a)}
     \put (18,52) {(b)}
     \put (59,52) {(c)}
     \put (18,25) {(d)}
     \put (59,25) {(e)}
    \end{overpic}
    \caption{(a) A schematic of the toy system we study for large network transport, comprising 102 dipoles in a helical pattern with detuning along its length. The colored regions correspond to thirds of the chain.  (b) The eigenstate localisation onto sites along the helix, and the size of the dots, represent how localised the particular eigenstate is on that site. The vertical axis represents the difference in the energy of the eigenstate and the ground state. (c) Dynamics of excitation moving through a toy model of a biased molecular spiral (a section of which is shown in the inset) with 102 sites solved using weak coupling Bloch-Redfield (dashed) and variational polaron master equation (thick). The populations are grouped into thirds of the chain length. (d)-(e) Plots showing the coherence length $L_\rho$ calculated using the variational polaron theory (black thick) and standard polaron theory (black dashed) varying (d) coupling to the vibrational modes (e) temperature of the bath. The minimum coherence length for fully localised sites ($V_{nm}=0$) is shown (black dotted). The red curves correspond to the mean (dashed) and standard deviation (dotted) of the variational parameters $\alpha_n$ along the chain. Critical values at which a stark localising transition occurs are depicted with a dotted grey line.}
    \label{fig:SpiralDyn}
\end{figure*}

\subsection{LH2 complex}
The next system we consider is the light-harvesting 2 (LH2) complex of \textit{Rhodopseudomonas molischianum}~\cite{Kundu2022}, shown in Fig.~\ref{fig:LH2Dyn}(a). This complex comprises 24 bacteriochlorophyll (BChl) molecules arranged in triplets: two lower-energy B850 species (aB850 and bB850) form the inner ring, and one higher-energy B800 species forms the outer ring. The triplet structure is illustrated in Fig.~\ref{fig:LH2Dyn}(b), where only the BChl molecules are shown, with B800 displayed in yellow and B850 in red. The system Hamiltonian was constructed using the values from ~\cite{Tretiak2000}. Here, the aB850 and B800 are resonant, and the bB850 is higher in energy by $80$~cm$^{-1}$. The strongest internal coupling is between adjacent aB850 and bB850 molecules with couplings of $408$ cm$^{-1}$ or $366$ cm$^{-1}$ depending on whether they are in the same or adjacent triplets. The outer B800s have the strongest coupling to their nearest aB850 and bB850 molecules with strengths of $52$ and $40$ cm$^{-1}$,  respectively. As such, we anticipate that the inner ring of B850 molecules will dimerise and excitations from the outer B800 ring will be transferred to the inner ring via FRET-like interactions.  
Each BChl couples to 50 normal-mode vibrations~\cite{Lh2Modes}. The normal modes spectral density takes the same form as that of the FMO given in Eqn.~\ref{eq:specnormal} with $\gamma=11~\text{cm}^{-1}$. In addition to the normal modes, we add a broad background spectral density to account for some features of environment inhomogeneity~\cite {Kundu2022}, which we model with a super-Ohmic form 
\begin{eqnarray}
    J_{\text{B}}(\omega) = A\frac{\omega^3}{\omega_c^3}\exp\{-\frac{\omega}{\omega_c}\},
\end{eqnarray}
with $A = 80~\text{cm}^{-1}$ and the cutoff frequency $\omega_c=100~\text{cm}^{-1}$.
The resulting spectral density is shown in Fig.~\ref{fig:LH2Dyn}(c), where normal modes have been grouped into clusters of five by colour. Experimentally, this system has been observed to transfer excitations from the outer B800 ring to the inner B850 ring on a $\sim 1$\,ps timescale at room temperature~\cite{Kundu2022}. Recent advances in numerically exact methods have made it possible to simulate such dynamics in full detail~\cite{Kundu2022}, though at serious computational cost. Using the variational master equation framework, however, we can explore these dynamics far more efficiently, due to the reduced complexity associated with master equation approaches.

We initialise the system with an excitation on one of the outer-ring B800 molecules. The resulting dynamics are shown in Fig.~\ref{fig:LH2Dyn}(d), and exhibit strong qualitative agreement with the numerically exact results of Ref.~\cite{Kundu2022}, even with slightly different system parameters. We observe a rapid transfer occurring over approximately $1.3$\,ps, during which over $80\%$ of the population accumulates within the inner B850 ring. In contrast, the weak-coupling dynamics significantly overestimate decay rates and underestimate transfer into neighbouring B800 sites.

Because simulations of networks of this size are computationally inexpensive in the variational framework, we can directly probe the influence of environmental structure on transport. In Fig.~\ref{fig:LH2Dyn}(c), the normal modes are grouped into coloured clusters containing five modes; by progressively adding these clusters to the environment, we can isolate the dynamical impact of the addition of specific sets of modes. To quantify transport, we define the parameter $\tau_{0.8}$ as the time required for $80\%$ of the exciton population to reach the central B850 ring from an initially excited B800 site.

Figure~\ref{fig:LH2Dyn}(e) shows how $\tau_{0.8}$ varies as the normal-mode clusters are introduced. Interestingly, the behaviour is highly non-monotonic, indicating that the addition of some vibrational modes hinders energy transfer while others enhance it. In particular, we find optimal transport when only the first half of the mode clusters are included. We anticipate that after this point, the vibrational modes become sufficiently high-frequency and off-resonant, which leads to a suppression of transfer rates. The ability to scan through individual environmental modes constitutes a powerful feature of the variational framework, enabling detailed study of how specific environmental degrees of freedom shape transport and suggesting new strategies for engineering environments that enhance exciton transfer in molecular systems.

\subsection{Toy model helix}
Finally, to demonstrate the scalability of our approach, we consider a helical model consisting of 102 sites and compute the dynamics using the scalable variational framework. The helix comprises 34 trimers, and a schematic of the network is shown in Fig.~\ref{fig:SpiralDyn}(a). In principle, larger systems can be simulated with this method; however, the 102-site case serves as a proof of principle, as eigenstates near the centre of the chain approach translational invariance, as seen in Fig.~\ref{fig:SpiralDyn}(b), suggesting that no new qualitative features are expected with increasing system size. The helical structure exhibits an energy gradient along its length, introducing a directional component to exciton transport reminiscent of phycobilisome architectures~\cite{Dodson2022}. This detuning is set to $40$ cm$^{-1}$ between adjacent triplets, and the dipole-dipole coupling is set such that nearest-neighbour couplings within a triplet are $350$ cm$^{-1}$. This leads to a delocalisation of its eigenstates as shown in Fig.~\ref{fig:SpiralDyn}(b), where we have taken the dipoles to be tangential to the helix path (see Appendix for full details of the Hamiltonian). The eigenstructure exhibits interesting features, with partial localisation of eigenstates onto sites; despite the linear detuning along the chain, we observe a strong non-linear dependence of the energy on the sites where the eigenstates are most localised. This is due to the detuning between triplets ($\approx 40~\text{cm}^{-1}$) being small compared to nearest neighbour coupling ($\approx 350~\text{cm}^{-1}$), as such, the dominant features are from triplet delocalisations and the detuning acts as a perturbation of this. We associate with each site a super-ohmic spectral density of the form
\begin{eqnarray}
    J_S(\omega) = A \frac{\omega^3}{\omega_c^3}\exp\{-\frac{\omega}{\omega_c}\},
\end{eqnarray}
with $A=180~\text{cm}^{-1}$ and cutoff frequency $\omega_c = 200~\text{cm}^{-1}$. 
The resulting dynamics are shown in Fig.~\ref{fig:SpiralDyn}(c), where we track excitations across a partitioning of the helix into three parts. In contrast to the previous examples, the variational master equation predicts significantly faster energy transfer through the network than the weak-coupling theory. This enhancement may arise from two mechanisms that depend sensitively on both the choice of spectral density and the structure of the Hamiltonian (and therefore the underlying molecular geometry). Due to the super-Ohmic form of the spectral density used here, we expect weak environmental coupling between near-resonant sites in the Markovian weak-coupling limit. Incorporating non-Markovian effects allows faster transitions. Moreover, because the variational optimisation can, in principle, yield distinct displacements for each site, we expect the corresponding reorganisation energies ($R_n$) to be strongly site dependent. This introduces local asymmetries into the system Hamiltonian, an effect known to enhance energy transport~\cite{Dubi2020,Coates2023}. Investigating the role of variationally induced asymmetries in facilitating enhanced transport will be the focus of future work.

To capture dynamics across such a large network, the simulations necessarily require long propagation times (on the order of thousands of picoseconds). This further illustrates the capability of the variational method not only to simulate large systems but also to evolve them over long timescales, enabling future studies that resolve the intrinsic timescales associated with optical and vibrational dynamics. Highly relevant for studying absorption and emission spectra of such systems. 

Additionally, the computational efficiencies of the partitioned variational parameter solver allow us to scan across different environmental configurations and examine their impact on delocalisation within the network. As a figure of merit, we use the coherence length of the thermal state
$\tilde{\rho}^\beta = {e^{-\beta \tilde{H}_0}}/{\mathrm{Tr}\!\left(e^{-\beta \tilde{H}_0}\right)}.$

\begin{equation}
    L_\rho =\frac{1}{N} \frac{(\sum_{ij}|\rho^\beta_{ij}|)^2}{\sum_{ij}(\rho^\beta_{ij})^2},
\end{equation}
where $\rho_\beta$ corresponds to the density matrix in the untransformed frame, which under a Born approximation relates to the density matrix in the variational frame via 
\begin{eqnarray}
    \rho^\beta_{ij} = \begin{cases}
        \tilde{\rho}^\beta_{ii}&, \hspace{5pt} i=j\\
        \tilde{\rho}^\beta_{ij}\mathcal{B}_{i}\mathcal{B}_{j}&, \hspace{5pt}i\neq j.
    \end{cases}
\end{eqnarray}
We now modify the coupling to the environment by varying the coupling strength through the parameter $A$ in the spectral density. The results are shown in Fig.~\ref{fig:SpiralDyn}(d), where a monotonic decrease in the coherence length is observed. When compared with the polaron theory result (dashed), we see a much slower drop-off in the coherence length. This suggests that polaron theory greatly overestimates the amount of localisation in the network.   We also plot the mean and standard deviation of the variational parameters along the spiral (in red), which qualitatively follow the same dependence on the coupling.

A unique and non-trivial feature emerges from this analysis: there is a sharp drop in the coherence length around a critical coupling value of $A \approx 320~\text{cm}^{-1}$, beyond which the system rapidly approaches the fully delocalised value for the network. This behaviour is a surprising prediction of the variational polaron theory, suggesting that at a critical level of environmental coupling, the system transitions from a regime where eigenstate delocalisation is possible (quantum regime) to one in which the eigenstates are fully localised and FRET-like dynamics dominate (classical regime). We note that, in the detuned spiral, the fully delocalised state does not saturate the coherence-length bound of $1/N$, but the value shown in Fig.~\ref{fig:SpiralDyn}(d) by the black dotted line. Furthermore, this sharp transition is not a property of the standard polaron theory, which shows a much smoother transition towards the delocalised state. 

In Fig.~\ref{fig:SpiralDyn}(e), we perform a similar analysis while increasing the temperature of the environment. Again, we observe a pronounced transition at a critical temperature $T_C \approx 600\,\text{K}$, to the fully delocalised state. The growth in the coherence length can be understood as due to the chain having a detuning along its length; higher temperatures enable more of the length of the chain to become thermally accessible, though this still represents a minimum after the transition, as shown by the black dotted line.

Another intriguing feature is associated with the standard deviation of the variational parameters $\alpha_n$. In both the temperature and coupling scaling cases, the standard deviation is not monotonic across the parameter regime and reaches a maximum roughly in the centre of the range. Reaching a maximum value indicates that the spread of $\alpha_n$s is highest or that there are more distinct $\alpha_n$s, and thus more distinct ${R}_n$ energy shifts at each site. This breaks the site degeneracy, introducing local energy barriers. Such energy barriers have been shown to enhance energy transport in similar systems~\cite{Dubi2020,Coates2023}. A full exploration of the intriguing features uncovered for this helix is left for future work. 


\section{Conclusion}
\label{sec:Conc}
Our framework unlocks efficient simulations of quantum transport in physical networks comprising hundreds of sites, regardless of environmental coupling strength. This is achieved with a variationally optimised polaron master equation approach that explicitly utilises the multi-scale nature of realistic networks. Specifically, our framework combines three key advances over the previous literature: (i) introducing a convergent local variational optimisation scheme, a method that scales efficiently while retaining the accuracy of the full Hilbert space variational optimisation, (ii) calculating a succinct form for the variational parameter $\alpha_n$,  and (iii) the development of analytic expressions for non-Markovian decay rates and the variational polaron propagator to enhance both computational efficiency and physical insight by introducing displaced Matsubara frequencies. Together, these advances enable the study of energy transport systems accounting for strong coupling effects in ways previously unattainable at unprecedented scales. 

We showcase the framework across a range of physically significant systems, including the canonical seven-site Fenna–Matthews–Olson complex benchmark found in green sulfur bacteria and the larger twenty-four-site LH2 complex from purple non-sulfur bacteria. We further demonstrate its scalability on a model system comprising more than one hundred sites, while consistently incorporating the influence of complex vibrational environments in each case. These applications reveal the role of environmental structure in governing transport efficiency and uncover a coupling- and temperature-driven localisation transition in transport-relevant eigenstates.

As a future perspective, our methodology could be developed using efficient representations of the Liouvillian operator, eliminating the current bottleneck in the memory cost for long-time propagation. This could be achieved through tensor network and reduced rank representations of the Liouvillian and the density matrix to compress the complexity of the Hilbert space by removing states associated with highly entangled spatially distant sites, or the deployment of cumulant expansions of the Liouvillian. 

\acknowledgements{
 A. Burgess and E.M. Gauger thank the Leverhulme Trust for support through grant number RPG-2022-335. 
 A. Burgess and E.M. Gauger also thank co-funding by Volkswagen Foundation - 0200195.
}

\bibliography{main}

\newpage 
\clearpage
\onecolumngrid
\appendix
\section*{Appendix}

\section{Master equation formulation in the variational frame}
\label{sec:master}
By writing the interaction Hamiltonian in the form $\tilde{H}_I = \sum_{i=1}^{N^2}S_i \otimes E_i$ (with interaction picture counterpart $\tilde{H}_I(t) = \sum_{i=1}^{N^2}S_i(t) \otimes E_i(t)$ ) we can rewrite the master equation in terms of system operators $S_i$ and two-time environmental correlation functions $\Lambda_{ij}(t-s) = \tr_E\left\{E_i(t)E_j(s) \rho_R\right\}$. In the Schr\"odinger picture, the master equation takes the form:
\begin{equation}
\frac{\partial \tilde{\rho}_S(t)}{\partial t} = - i [\tilde{H}_0,\tilde{\rho}_S(t)]  
- \sum_{i,j} \int_0^td s \bigg(\Lambda_{ij}(s)\{S_i S_j(s) \tilde{\rho}_S(t) - S_j(s) \tilde{\rho}_S(t) S_i\} +  \text{h.c.}\bigg),\label{eq:me}
\end{equation}
where h.c. correspond to the Hermitian conjugate. 
The interaction Hamiltonian system operators 
can be split into three distinct groups in the following way:
\begin{equation}
S_i = 
\begin{cases}
\ketbra{n}{n} = S^z_n \quad &1 \leq i\leq N,\\
\ketbra{n}{m} + \ketbra{m}{n} = S^x_{nm} \quad &N <i\leq \frac{1}{2}N(N+1),\\
i\ketbra{n}{m} - i\ketbra{m}{n} = S^y_{nm} \quad  &\frac{1}{2}N(N+1) <i\leq N^2,
\end{cases} .
\end{equation}
which leads in turn to the generation of three varieties of non-zero time correlation function~\cite{pollock2013multi,Rouse2024Polariton}. The first type is due to the linear interaction term, 
$\tilde{H}_L$, and is therefore of the same form as those that appear in the standard weak coupling master equation:
\begin{align}
&\Lambda^{zz}_{n}(t) = \phi^{zz}_n(t)   \int_0^\infty{d\omega \; J_n(\omega)\left[1-F_n(\omega)\right]^2\left[\cos(\omega t) \coth(\beta \omega /2) - i \sin(\omega t)\right]}, \label{eq:phizz}
\end{align}
where $F_n(\omega)$ is the continuum version of the optimised displacement function. The second type arises from the displacement operator interaction, $\tilde{H}_D$, and is the only type to appear in the fully displaced polaron master equation:
\begin{align}
\Lambda^{xx}_{nmpq}(t) = &\frac{1}{2}V_{nm}V_{pq}\mathcal{B}_n \mathcal{B}_m \mathcal{B}_p \mathcal{B}_q  \bigg\{ \exp\left[\delta_{np} \phi^{xy}_n(t) +\delta_{mq} \phi^{xy}_m(t) \right] + \exp \left[-\delta_{np} \phi^{xy}_n(t) -\delta_{mq} \phi^{xy}_m(t) \right]-2\bigg\},
\end{align}
\begin{align}
&\Lambda^{yy}_{nmpq}(t) = \frac{1}{2}V_{nm}V_{pq}\mathcal{B}_n \mathcal{B}_m \mathcal{B}_p \mathcal{B}_q \bigg\{ \exp\big[\delta_{np} \phi^{xy}_n(t) +(\delta_{mq}-\delta_{mp}) \phi^{xy}_m(t) \big] \nonumber- \exp\left[-\delta_{np}\phi^{xy}_n(t) -(\delta_{mq}-\delta_{mp})\phi^{xy}_m(t) \right]\bigg\},
\end{align}
where
\begin{eqnarray}
\phi^{xy}_n(t) = \int_0^\infty{d\omega \; \frac{J_n(\omega)}{\omega^2} F_n(\omega)^2 \left[\cos(\omega t) \coth(\beta \omega /2) - i \sin(\omega t)\right]},\label{eq:phixy}
\end{eqnarray}
and the $\delta_{nm}$ are Kronecker deltas. Finally, the third type appears in the more general variational master equation due to an overlap between the two types of interaction:
\begin{equation}
\Lambda^{yz}_{nmp}(t) = \delta_{np} V_{nm}\mathcal{B}_n \mathcal{B}_m \phi^{yz}_n(t),
\end{equation}
with
\begin{align}
\phi^{yz}_n(t) = \int_0^\infty d\omega & \; \frac{J_n(\omega)}{\omega} F_n(\omega)\left[1-F_n(\omega)\right] \big[\sin(\omega t) \coth(\beta \omega /2) + i \cos(\omega t)\big].\label{eq:phiyz}
\end{align}

\section{Non-Markovian Decay Rates}
\label{app:Rates}
One of the remaining key challenges in implementing the full non-Markovian rates for the variational polaron master equation is associated with taking Fourier transforms of the correlation functions $\Lambda_{ij}(s)$, particularly when accounting for the variational distributions $F_n$. This is particularly the case for the exponential of the polaron propagator that appears in $\Lambda_{nmpq}^{xx,yy}$ (see Appendix~\ref{sec:master}. Here, we derive an analytical form for these rates to enhance both the accuracy and efficiency of their calculation.

We will start by considering the variational polaron propagator term 
\begin{equation}
    \phi(t) = \int^\infty_0 d\omega \frac{J(\omega)}{\omega^2}F(\omega)^2\left[\cos(\omega t)\coth(\beta \omega/2)-i\sin(\omega t) \right],
\end{equation}
where, for convenience, we have dropped the explicit indexing. 
The form of the variational distribution is 
\begin{equation}
    F(\omega) = \frac{\omega}{\omega+ \alpha \coth(\beta\omega/2)}.
\end{equation}
We aim to utilise a method of residues and thus must consider the poles in the complex plane of the integrand in $\phi(t)$, these poles can originate in four places, the first is the spectral density $J(\omega)$, the second is in $F(\omega)$, the third in $\coth(\beta\omega/2)$ and the final in $\frac{1}{\omega^2}$. We note that, however, $F^2(\omega)$ removes all the poles associated with the latter two functions, and as such, we need only consider the poles of $J(\omega)$ and $F(\omega)$. To determine the poles of the variational distribution, we consider the Matsubara decomposition of the hyperbolic cotangent function 
\begin{equation}
     \text{coth}\left(\frac{\beta\omega}{2}\right) = 2T\left(\frac{1}{\omega}+ \sum_{k=1}^\infty \frac{2\omega}{\omega^2 +(2\pi k T)^2} \right),
\end{equation}
and thus 
\begin{equation}
    F(\omega) = \frac{\omega}{\omega+\alpha(2T\left(\frac{1}{\omega}+ \sum_{k=1}^\infty \frac{2\omega}{\omega^2 +(2\pi k T)^2} \right))},
\end{equation}
truncating the series up to some finite value $N_m$ allows for the rewriting of the $F(\omega)$ as a complex rational function of degree $2N_m+2$ in both numerator and denominator with 
\begin{equation}
    F(\omega) = \prod_{k=0}^{N_m}\frac{(\omega^2+(2\pi kT)^2)}{(\omega^2-x_k^2)},
\end{equation}
where $\pm x_k$ are the roots of the equation $\omega+\alpha\coth(\beta \omega/2)$, which can be approximated via inversion of the Taylor series about the roots of $\coth(\beta \omega/2)$, at the Matsubara frequencies $\omega =2i\pi k T$. This leads to a collection of shifted Matsubara frequencies, brought about by the partial displacement, as shown in Fig.~\ref{fig:AnaRates}(a).  See Appendix \ref{app:Poles} for more details on these roots. Inside of $\phi(t)$, this $F(\omega)$ is squared as such, we have second-order poles about $\pm x_n$. We now take the form of the spectral density to be Ohmic Drude-Lorentz 
\begin{equation}
    J(\omega) = \frac{\lambda \gamma \omega}{(\omega-\omega_c)^2 + \gamma^2},
\end{equation}
we choose such a form as we can typically decompose any spectral density into a sum of such spectral densities. We will also, for clarity of analysis, assume $\omega_c=0$, this is not necessary, but cleans up the following derivation. Such a spectral density has poles at $\omega=  \pm i \gamma$. We also introduce the mode memory kernel term 
\begin{equation}
    p(\omega,t) = \coth(\beta\omega/2)\cos(\omega t)-i\sin(\omega t),
\end{equation}
such that the polaron propagator can be written as 
\begin{equation}
    \phi(t) = \int^\infty_0 d\omega \frac{J(\omega)}{\omega^2}F^2(\omega)p(\omega,t),
\end{equation}
noting that the integrand is an even function of $\omega$ allows for the rewriting 
\begin{equation}
     \phi(t) = \frac{1}{2}\int^\infty_{-\infty} d\omega \frac{J(\omega)}{\omega^2}F^2(\omega)p(\omega,t).
\end{equation}

From this, we can use the residue theorem to determine the polaron propagator to be in the form 
\begin{equation}
    \phi(t) = \sum_{j=-1}^{N_m}(a_j + b_jt)e^{-\gamma_jt},
\end{equation}
where the exact values of $a_j,b_j$ and $\gamma_j$ are shown in Appendix \ref{app:ExpForm}. 
To validate that this procedure faithfully captures the polaron propagator in the variational frame, we demonstrate the convergence of the analytics for a specific case in which we are in a low-temperature regime. The results of this are shown in Fig.\ref{fig:AnaRates}(b), where we see a perfect matching between analytic and numerical approaches to calculating the polaron propagator. 


\begin{figure*}[t]
    \centering
    \begin{overpic}[width=\linewidth]{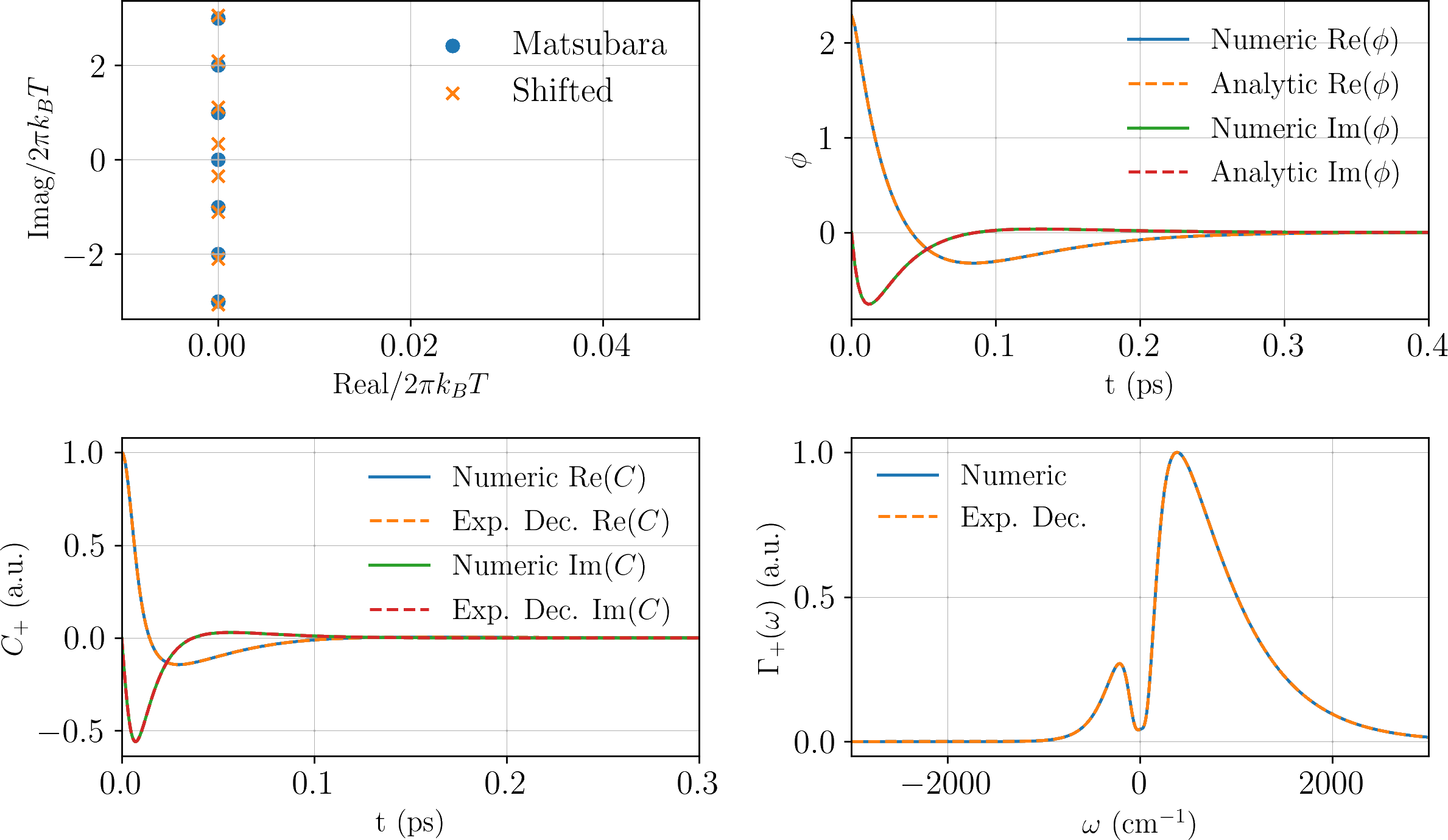}
     \put (0,57) {(a)}
     
     \put (51,57) {(b)}
     \put (0,27) {(c)}
     \put (51,27) {(d)}
    \end{overpic}
    \caption{(a) Plot of the complex plane with the Matsubara frequencies and the `shifted' Matsubara frequencies generated by the variational distribution $F(\omega)$. At $T=300K$ and $\alpha =0.1$eV. (b) Calculation of the polaron propagator using analytic (crosses) and numerical (thick) results for an Ohmic Lorentz-Drude spectral density with $\gamma = 0.1$ eV, $\alpha= 0.005$ eV, $\lambda =0.1$ eV and $T=300$ K. (c)-(d) Calculation of polaron-variational frame correlation function (c)  and rate function (d) from an exponential-type decomposition. For a super-Ohmic spectral density approximation of the multi-peaked LH2 spectral density with parameters for the super-Ohmic spectral density of a cut-off frequency $\omega_c=300~\text{cm}^{-1}$ and $\lambda = 0.628$ and variational parameter $\alpha = 60~ \text{cm}^{-1}$ at $T=300~K$.}
    \label{fig:AnaRates}
\end{figure*}
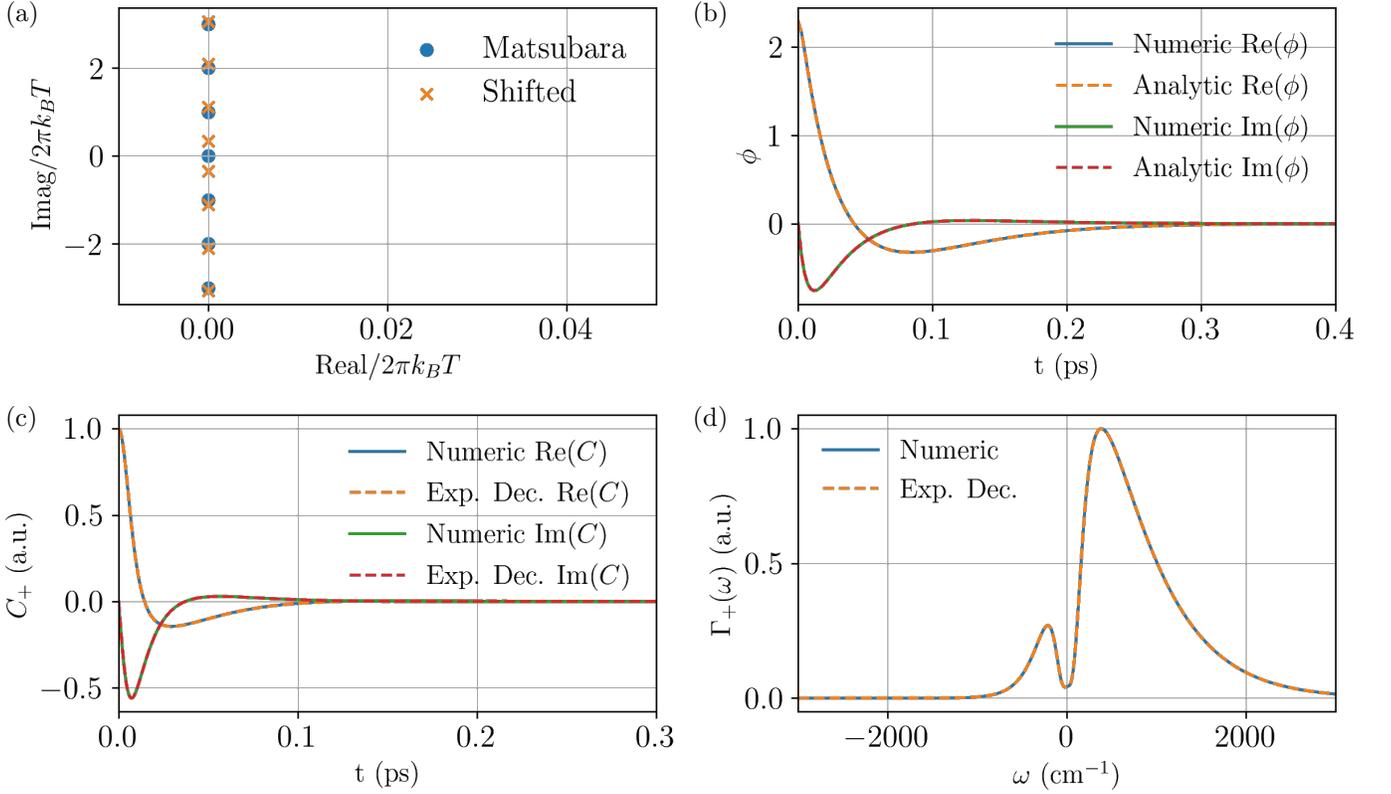



Having found an exact form of the polaron propagator, we can motivate the study of when the polaron propagator takes a simple exponential form. This is useful as for some low-temperature systems, the number of Matsubara frequencies required to reach convergence can be large, and as such, one can instead numerically fit the polaron propagator with significantly fewer exponential functions, reducing overheads. 
Such a polaron propagator would take the form

\begin{equation}
    \phi(t) =  \sum_{i=1}^N c_i e^{-\gamma_i t}, \hspace{6pt} c_i,\gamma_i \in \mathbb{C}.
\end{equation}

The correlation functions that are particularly complex to calculate are of the form  
\begin{equation}
    C_\pm(t) = e^{\pm\phi(t)}-1. 
\end{equation}
We start by performing a Taylor series in $\phi$ of the correlation function
\begin{equation}
    C_\pm(t) = k\sum_{n=1}^\infty \frac{[\pm\phi(t)]^n}{n!}.
\end{equation}
This is in part due to the frequency domain integral in the exponential function, making calculations of frequency-resolved features challenging.
Substituting the exponential form of the phonon propagator into the correlation functions yields
\begin{gather}
    C_\pm(t) = \sum_{n=1}^\infty (\pm1)^n \sum_{\mathbf{l}\in \mathbb{P}_n^N}\prod_i \frac{c_i^{l_i}e^{-l_i\gamma_i t}}{l_i!},\\
    = \sum_{n=1}^\infty (\pm1)^n \sum_{\mathbf{l}\in \mathbb{P}_n^N}F_\mathbf{l}e^{-\Gamma_\mathbf{l} t}.
\end{gather}
where $\mathbb{P}_n^N \subset \mathbb{N}^N$ is the set of integer partitions of length $N$ such that the sum of the elements add to $n$, i.e, 
\begin{equation}
    \mathbf{l}\in \mathbb{P}_n^N \implies \sum_{i=1}^N l_i = n,
\end{equation}
and we have defined 
\begin{gather}
    F_\mathbf{l} =\prod_i\frac{c_i^{l_i}}{{l_i!}} ,\\
    \Gamma_\mathbf{l} = \sum_i l_i\gamma_i.
\end{gather}
This form may not appear much simpler than before. However, we can take advantage of the fact that $|\phi|$ is typically of order unity and thus the Taylor series can be quickly truncated. Furthermore, when we consider the non-Markovian rates associated with these correlations, they become almost trivial to calculate
\begin{align}
    \Gamma_\pm(\omega,t) &= \int^t_0 dt' e^{i\omega t'} C_\pm(t') \\
    &= \sum_{n=1}^\infty (\pm1)^n \sum_{\mathbf{l}\in \mathbb{P}_n^N}\frac{F_\mathbf{l}
}{\Gamma_\mathbf{l} -i\omega}(1-e^{i\omega t - \Gamma_\mathbf{l}t}).
\end{align}
Now that we have developed the framework for calculating the non-Markovian rates, we apply it directly to a spectral density of the form 
\begin{equation}
    J(\omega) = \lambda \frac{\omega^3}{\omega_c^2}\exp\{-(\omega/\omega_c)\}.
\end{equation}
The results of this are shown in Fig.~\ref{fig:AnaRates}(c) and (d), where only six exponentials for both the real and imaginary components of $\phi(t)$ are required to recover the correlation function $C_+(t)$ as well as the entire rate function $\Gamma_+(\omega) = \Gamma_+(\omega,\infty)$. We can note that this procedure of performing exponential fitting greatly reduces the required number of exponentials compared to the analytic calculations shown previously and expands the domain of possible spectral densities beyond sums of Lorentzians. Furthermore, this procedure is less prone to erroneous resurgences that can occur in finite domain numerical Fourier transforms.  The parameters for the spectral density are a super Ohmic approximation of the LH2 spectral density with a cut-off frequency $\omega_c=300~\text{cm}^{-1}$ and $\lambda = 0.628$, which leads to an environment reorganisation energy of 377 $\text{cm}^{-1}$.

\section{Parameters used for large helical energy transfer networks.}
\label{app:Params}
For the large toy-model spiral system, we have positions along a spiral such that the Cartesian coordinates of the sites are given by
\begin{equation}
    (x_{ij},y_{ij},z_{ij}) =  (r\cos(i\theta ),iv+js,r\sin(i\theta)),
\end{equation}
with $\theta=0.6$, $r=4$nm, $v=1$nm, $s=0.8$nm, and there are 34 triplets such that $i\in\{0,..,33\}$ and $j\in\{0,1,2\}$. We associate dipoles $\mathbf{d}_n$ with each site that are tangential to the spiral, such that the dipoles couple to each other via the dipole-dipole coupling
\begin{eqnarray}
    V_{nm} =  \frac{\big(\mathbf{d}_n\cdot \mathbf{d}_m -3 (\mathbf{d}_n\cdot \mathbf{\hat{r}}_{nm})(\mathbf{d}_m\cdot \mathbf{\hat{r}}_{nm})\big)}{4\pi\epsilon_0 r_{nm}^3},
\end{eqnarray}
where $\mathbf{r}_{nm}$ is the relative separation between the two sites. The dipoles are normalised such that for parallel dipoles that are perpendicular to a separation of 1~nm, couple with a strength of $0.4$~eV. Using the joint index $n=3i+j$, the system Hamiltonian can be written as 
\begin{eqnarray}
    H_0 = \sum_{n}(\epsilon+i\delta )\ketbra{n}{n} + \sum_{n\neq m} V_{nm} \ketbra{n}{m},
\end{eqnarray}
where the first site energy is $\epsilon=2$ eV and the relative detuning between triplets is $\delta=0.005$ eV, introducing an energy gradient along the spiral. The spiral model had the associated super-Ohmic spectral density
\begin{eqnarray}
    J_S(\omega) = A \frac{\omega^3}{\omega_c^3}\exp\{-\frac{\omega}{\omega_c}\},
\end{eqnarray}
with $A=180~\text{cm}^{-1}$ and cutoff frequency $\omega_c = 200~\text{cm}^{-1}$. For the larger network shown in the optimisation section with 3000 sites, we use the same system allowing $i\in\{0,2999\}$.
\section{Poles of partial displacement function}
\label{app:Poles}
To utilise the residue theorem to calculate the correlation functions described above, we need to find the poles of the variational distribution function $F(\omega)$. This can be achieved by approximating the solutions via a displaced Taylor series

\begin{align}
    x_n = &i \frac{\bigg(-128 \pi ^4 \alpha ^2 \beta ^2 n^4 (\alpha  \beta +6)-15 \alpha ^4 \beta ^4 (\alpha  \beta +4)^3+3072 \pi ^8 n^8 +1536 \pi ^6 \alpha  \beta  n^6 +4 \pi ^2 \alpha ^3 \beta ^3 n^2 \left(3 \alpha ^2 \beta ^2+64 \alpha  \beta +192\right)\bigg)}{(1536 \pi ^7 \beta  n^7)},
\end{align}
This allows for the writing of the variational distribution as 
\begin{equation}
F(\omega) = \prod_{n=0}^N\frac{(\omega^2+(2n\pi T)^2)}{(\omega^2+|x_n|^2)}.
\end{equation}
One can also note that the variational distribution removes the poles from $\coth(\beta\omega/2)$ associated with the Matsubara frequencies and introduces `shifted' Matsubara frequencies. We show these in Fig.\ref{fig:AnaRates}(a), where we can see the splitting of the zero frequency Matsubara frequency into two distinct modes.

\section{Exponential form for polaron propagator}
\label{app:ExpForm}
In the main text, we determined that the polaron propagator can be expanded in a series of exponentials as 
\begin{equation}
    \phi(t) = \sum_{j=-1}^{N_m}(a_j + b_jt)e^{-\gamma_jt}
\end{equation}
The coefficients are given by 
\begin{equation}
    \gamma_j = \begin{cases}
        \gamma,\hspace{12pt} j=-1,\\
        |x_j|, \hspace{3pt} j\geq 0,
    \end{cases}
\end{equation}

\begin{eqnarray}
    a_j = \begin{cases}
        -i\pi\frac{\lambda}{2\gamma}F(i\gamma)^2\Tilde{p}(i\gamma,0),\hspace{40pt} &j=-1,\\
        i\pi\big(J'(x_j)\Tilde{F}_j(x_j)^2\Tilde p(x_j,0) +2J(x_j)\Tilde{F}_j'(x_j)\Tilde{F}_j(x_j)\Tilde{p}(x_j,0)  ,+J(x_j)\Tilde{F}_j(x_j)^2\Tilde p'(x_j,0)\big) \hspace{8pt} &j\geq 0,
    \end{cases}
\end{eqnarray}
with 
\begin{equation}
    \Tilde{p}(\omega,t) = \coth(\beta\omega/2)e^{i\omega t}-e^{i\omega t}
\end{equation}
and 
\begin{equation}
    \Tilde{F}_j(\omega) = \frac{F(\omega)(\omega-x_j)}{\omega},
\end{equation}
and 
\begin{equation}
    b_j= \begin{cases}
        0, j=-1,\\
        -\pi\left(J(x_j)\Tilde{F}_j(x_j)^2(\coth(\beta x_j/2) -1)\right),\hspace{3pt} j\geq 0,
    \end{cases}
\end{equation}
where we have defined $x_j$ as the poles associated with the displacement function outlined in the next section.

\section{Coherence length calculation}
The coherence length was used as a figure of merit for delocalisation in the biased spiral system. We calculate here the coherence length for the fully localised configuration. In the high phonon coupling limit, we expect all off-diagonal terms in the Gibbs state to be zero and all diagonal terms to be (using the joint index notation $n=3i+j$) 
$\rho_{nn}\propto e^{-\beta i \delta}$. Thus, the coherence length defined by 
\begin{equation}
    L_\rho = \frac{1}{N} \frac{(\sum_n \rho_{nn})^2}{\sum_n \rho_{nn}^2},
\end{equation}
can be found using geometric series and noting that there are $34$ triplets with equal energy to be 
\begin{equation}
    L_\rho = \frac{1}{34}\frac{(1-r^{34})(1+r)}{(1+r^{34})(1-r)}\approx \frac{1}{34}\frac{1+r}{1-r}\approx 0.304,
\end{equation}
where $r = e^{-\beta\delta}$, thus this creates the lower bound for the coherence length in this system.

\end{document}